\documentclass[sigconf, screen]{acmart}

\acmSubmissionID{377}

\usepackage{hyperref}
\usepackage{latexsym}
\usepackage{mathrsfs}
\usepackage{multirow}
\usepackage{amsmath}
\usepackage{amsfonts}
\usepackage{subfigure}
\usepackage{graphicx}
\usepackage{tabularx}
\usepackage[boxed,ruled,lined]{algorithm2e}
\usepackage{algorithmic}
\usepackage{comment}
\usepackage{balance}
\usepackage{bm}
\usepackage[inline]{enumitem}
\usepackage[skip=1mm]{caption}

\usepackage{acronym}
\newtheorem{theorem}{Theorem}

\newcommand{\ie}{\emph{i.e., }}

\newcommand{\wrt}{\emph{w.r.t. }}

\acrodef{IR}{information retrieval}

\usepackage{amsmath,amsfonts,bm}

\def\eqref#1{equation~\ref{#1}}

\def\1{\bm{1}}

\DeclareMathAlphabet{\mathsfit}{\encodingdefault}{\sfdefault}{m}{sl}
\SetMathAlphabet{\mathsfit}{bold}{\encodingdefault}{\sfdefault}{bx}{n}

\DeclareMathOperator*{\argmax}{arg\,max}

\author{Chen Xu}
\orcid{https://orcid.org/0000-0002-3070-9358}
\affiliation{%
  \institution{\mbox{Gaoling School of Artificial Intelligence}\\Renmin University of China}
  \city{Beijing}
  \country{China}
}
\email{xc\_chen@ruc.edu.cn}

\author{Jujia Zhao}
\orcid{https://orcid.org/0009-0003-6951-7593}
\affiliation{
  \institution{Leiden University}
  \city{Leiden}
  \country{The Netherlands}
}
\email{j.zhao@liacs.leidenuniv.nl}

\author{Wenjie Wang}
\orcid{https://orcid.org/0000-0002-5199-1428}
\affiliation{%
   \institution{University of Science and Technology of China}
   \city{Hefei}
   \country{China}
}
\email{wenjiewang96@gmail.com}

\author{Liang Pang}
\orcid{https://orcid.org/0000-0003-1161-8546}
\affiliation{%
  \institution{Institute of Computing Technology Chinese Academy of Sciences}
  \city{Beijing}
  \country{China}
}
\email{pangliang@ict.ac.cn}

\author{Jun Xu}
\orcid{https://orcid.org/0000-0001-7170-111X}
\authornote{Jun Xu is the corresponding author. }
\affiliation{%
    \institution{\mbox{Gaoling School of Artificial Intelligence}\\Renmin University of China}
  \city{Beijing}
  \country{China}
}
\email{junxu@ruc.edu.cn}

\author{Tat-Seng Chua}
\orcid{https://orcid.org/0000-0001-6097-7807}
\affiliation{
  \institution{National University of Singapore}
  \country{Singapore}
}
\email{chuats@comp.nus.edu.sg}

\author{Maarten de Rijke}
\orcid{https://orcid.org/0000-0002-1086-0202}
\affiliation{
  \institution{University of Amsterdam}
  \city{Amsterdam}
  \country{The Netherlands}
}  
\email{m.derijke@uva.nl}

\renewcommand{\shortauthors}{Chen Xu et al.}

\copyrightyear{2025}
\acmYear{2025}
\setcopyright{acmlicensed}\acmConference[SIGIR '25]{Proceedings of the 48th
International ACM SIGIR Conference on Research and Development in
Information Retrieval}{July 13--18, 2025}{Padua, Italy}
\acmBooktitle{Proceedings of the 48th International ACM SIGIR Conference on
Research and Development in Information Retrieval (SIGIR '25), July 13--18,
2025, Padua, Italy}
\acmDOI{10.1145/3726302.3730106}
\acmISBN{979-8-4007-1592-1/2025/07}



\title{The Art of Balancing Accuracy and Fairness: \\An Economic Perspective on Fair Re-Ranking}
\title{Two Sides of One Economic Coin: 
\\The Art of Balancing Accuracy and Fairness in Re-ranking}
\title{Two Sides of the Taxation Coin: \\The Art of Balancing Accuracy-Fairness in Fair Re-ranking}
\title{The Art of Accuracy-Fairness Trade-off in Re-ranking: \\ A Coupling of Income and Commodity Taxes}
\title{Elastic Fairness: An Economic Perspective on Fair Re-ranking}
\title{Reframing Fair Re-ranking Through Elasticity in Economics}
\title{Reframing Fair Re-ranking Through Tax Elasticity}
\title{Balancing Accuracy-Fairness in Re-ranking through Tax Elasticity}
\title{Unveiling Accuracy-Fairness in Re-ranking Through Elasticity}
\title[Understanding Accuracy-Fairness Trade-offs in Re-ranking through Elasticity]{Understanding Accuracy-Fairness Trade-offs\\ in Re-ranking through Elasticity}
\title[Understanding Accuracy-Fairness Trade-offs in Re-ranking through Elasticity in Economics]{Understanding Accuracy-Fairness Trade-offs in Re-ranking \\through Elasticity in Economics}

\begin{document}

\begin{abstract}
Fairness is an increasingly important factor in re-ranking tasks. 
Prior work has identified a trade-off between ranking accuracy and item fairness. 
However, the underlying mechanisms are still not fully understood.
An analogy can be drawn between re-ranking and the dynamics of economic transactions. 
The accuracy-fairness trade-off parallels the coupling of the commodity tax transfer process. 
Fairness considerations in re-ranking, similar to a commodity tax on suppliers, ultimately translate into a cost passed on to consumers.
Analogously, item-side fairness constraints result in a decline in user-side accuracy.
In economics, the extent to which commodity tax on the supplier (item fairness) transfers to commodity tax on users (accuracy loss) is formalized using the notion of elasticity.  
The re-ranking fairness-accuracy trade-off is similarly governed by the elasticity of utility between item groups.
This insight underscores the limitations of current fair re-ranking evaluations, which often rely solely on a single fairness metric, hindering comprehensive assessment of fair re-ranking algorithms.

Centered around the concept of elasticity, this work presents two significant contributions.
We introduce the Elastic Fairness Curve (EF-Curve) as an evaluation framework. This framework enables a comparative analysis of algorithm performance across different elasticity levels, facilitating the selection of the most suitable approach.
Furthermore, we propose ElasticRank, a fair re-ranking algorithm that employs elasticity calculations to adjust inter-item distances within a curved space.
Experiments on three widely used ranking datasets demonstrate its effectiveness and efficiency.

\end{abstract}

\maketitle

\section{Introduction}

Over the past decade, fairness has become an increasingly important and urgent topic on the \ac{IR} agenda~\cite{li2022fairness, deldjoo2022survey}.
Previous work proposes diverse fairness objectives to ensure a healthy ecosystem from economic or social perspectives~\cite {TaxRank, fairrec, xu2023p}.
However, fair re-ranking often entails a trade-off between ranking accuracy and item fairness, where improving one typically comes at the expense of the other~\cite{li2022fairness, deldjoo2022survey, TaxRank}. 
Various methods~\cite{xu2023p, fairrec, cpfair} have been proposed to mitigate this trade-off, but the underlying mechanisms of this phenomenon remain insufficiently understood.

\smallskip\noindent%
\textbf{An economic perspective on accuracy-fairness trade-off.}
The accuracy-fairness trade-off mirrors the coupling relationship of the commodity tax transfer process~\cite{ramsey1927contribution}. 
In economics, promoting fairness is achieved through taxation~\cite{nerre2001concept}. When a commodity tax is imposed on a supplier, the tax rate is not entirely absorbed by the supplier but is partially transferred to consumers.
When we relate re-ranking to economic transactions, we can view users as suppliers, item groups as consumers, and the ranking scores as the price; a more detailed correspondence can be seen in Table~\ref{tab:compare}. 
In this way, fairness in re-ranking functions like a commodity tax on the item side~\cite{TaxRank}, where the ranking scores adjusted by the fairness function (commodity tax for the supplier) transform into the cost of accuracy loss (commodity tax transferred to users). 
We provide a more detailed theoretical analysis of this analogy in Section~\ref{sec:method}.


\begin{figure}[t]  
    \centering    
    \includegraphics[width=0.95\linewidth]{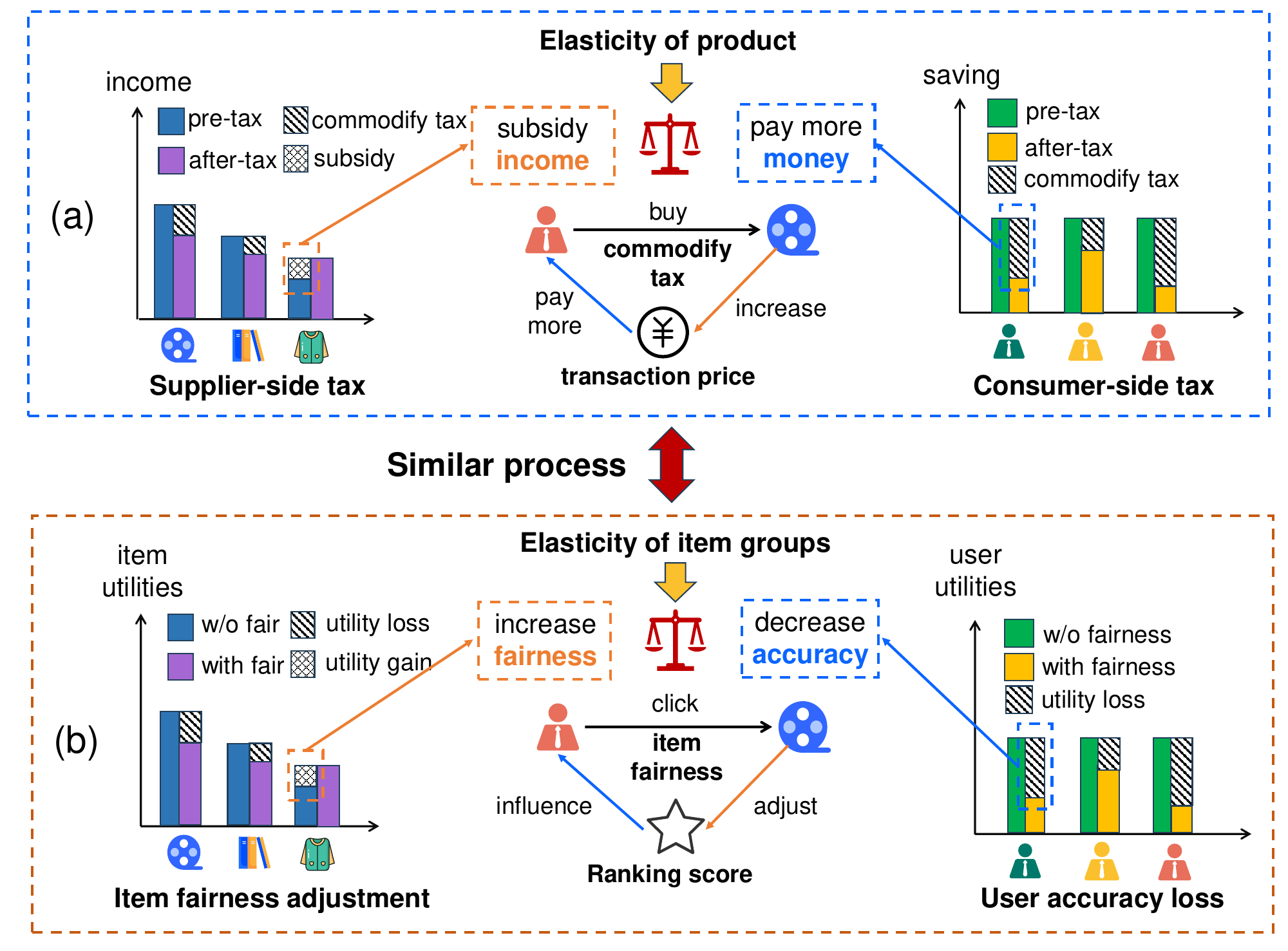}
    \caption{Parallels between (a) the commodity tax transfer process and (b) the accuracy-fairness trade-off in re-ranking.}
    \label{fig:intro}
\end{figure}

We use the example in Figure~\ref{fig:intro} to visualize this insight. 
In the economic transactions shown in Figure~\ref{fig:intro} (a), the imposition of a commodity tax on the supplier triggers a partial transfer of this tax burden to consumers in the form of a higher commodity price (a detailed example is in Section~\ref{sec:elastic_theory}).
Similarly, in Figure~\ref{fig:intro} (b), users will be exposed to certain items with high-ranking scores. When promoting fairness among items, poor item groups are assigned with higher ranking scores~\cite{TaxRank}, which alters the utility for some users. Consequently, part of the fairness cost is shifted to the users as an accuracy loss. 
Given the parallels between the two processes, we hypothesize that tools designed for analyzing taxation can be applied to understand accuracy-fairness trade-offs in re-ranking.

\smallskip\noindent%
\textbf{Understanding re-ranking through elasticity.}
In economics, elasticity theory~\cite{robbins1997elasticity} provides a framework for analyzing the transfer of commodity tax burdens to consumers, with the degree of transfer dependent on the elasticity of goods. Drawing an analogy, we demonstrate in Section~\ref{sec:evaulation} that the fairness-accuracy trade-off in re-ranking is similarly governed by the elasticity of utility between item groups. This insight reveals a key limitation of current fair re-ranking evaluations: their reliance on single fairness metrics, which effectively assess tax transfers under only specific elasticity assumptions, thereby hindering a comprehensive assessment of fair re-ranking algorithms.

Grounded in elasticity theory, we introduce the \emph{Elastic Fairness Curve} (EF-Curve), a framework for evaluating fair re-ranking algorithms. 
The EF-Curve visualizes algorithm performance across a spectrum of elasticities, with each point representing performance under a specific fairness metric. Intuitively, the EF-Curve illustrates that different fairness metrics measure the degree of support for item groups with varying levels of elasticity.
This framework facilitates a comparative analysis of algorithms, enabling the selection of the most suitable approach for diverse deployment scenarios. 
Moreover, the area enclosed by the EF-Curve and the axes, termed EF, provides a quantitative measure of overall algorithm performance.  

Furthermore, we introduce \emph{ElasticRank}, a fair re-ranking algorithm designed to optimize the EF metric. 
Grounded in elasticity theory, ElasticRank models the re-ranking space as curved, where inter-group distances are dynamically adjusted based on their respective elasticities. 
This approach intuitively prioritizes fairness for item groups with higher elasticity, minimizing the associated accuracy loss. 
Importantly, ElasticRank achieves this with comparable complexity as standard ranking algorithms.


\smallskip\noindent%
\textbf{Main contributions.}
We summarize the major contributions:
\begin{enumerate}[leftmargin=*]
    \item This research frames the re-ranking accuracy-fairness trade-off as a commodity taxation transfer problem. By employing elasticity theory from economics, we elucidate the intricate relationship between these competing objectives.

    \item Inspired by elasticity theory, we propose a novel evaluation framework for fair re-ranking algorithms, the EF-Curve. This framework facilitates comprehensive comparisons between different algorithms by visualizing their performance across a spectrum of fairness constraints.

    \item To optimize EF-Curve, we introduce ElasticRank, a novel fair re-ranking algorithm designed to optimize the EF-Curve. Rigorous empirical evaluation on three publicly available ranking datasets demonstrates that ElasticRank consistently surpasses state-of-the-art baselines.
\end{enumerate}

\section{Related Work}

\noindent%
\textbf{Fair re-ranking.}
Over the past decade, work on fair ranking tasks has rapidly grown in volume, driven by the need for a responsible and trustworthy ecosystem~\cite{lifairness, lipani2016fairness, deldjoo2022survey, xu2025fairdiversecomprehensivetoolkitfair}. 
Previous research often categorizes fair-aware methods into three categories based on ranking phases: pre-processing~\cite{Calmon17, xiong2024fairwasp}, in-processing~\cite{Tang23FairBias}, and post-processing (\ie re-ranking tasks)~\cite{xu2023p, fairrec}. The re-ranking phase is regarded as the most easily adaptable and practical stage in optimizing ranking systems~\cite{fairrec}.
During the re-ranking phase, the concept of fairness in re-ranking depends on the stakeholders involved~\cite{abdollahpouri2020multistakeholder, abdollahpouri2019multi}. 
Prior work has examined user-oriented fairness~\cite{abdollahpouri2019unfairness, li2021user} and item-oriented fairness~\cite{fairrec, xu2023p, TaxRank, singh2019policy, jaenich2024fairness, TaoSIGIRAP}. 
In this paper, we focus on item group fairness in re-ranking tasks. 

\smallskip\noindent%
\textbf{Metrics and algorithms in fair re-ranking.}
Fairness metrics vary widely across works, with different studies optimizing distinct metrics. For instance, some work~\cite{fairrec, TaoSIGIRAP} employs proportional fairness, \citet{do2022optimizing} focuses on the Gini Index, other work~\cite{xu2023p} prioritizes MMF, and TaxRank~\cite{TaxRank} optimizes $\alpha$-fairness. However, these approaches rely on single fairness metrics, which limits their ability to provide a comprehensive evaluation.

Previous work on re-ranking methods to improve item fairness can be divided into 
\begin{enumerate*}[label=(\roman*)]
\item regularized methods, which use a multi-task optimization approach with a linear combination of accuracy and fairness loss functions, incorporating a trade-off coefficient $\lambda$~\cite{xu2023p, do2022optimizing, cpfair}, and
\item constraint-based methods, which formulate the task as a constrained optimization problem to ensure that fairness metrics do not exceed a specified threshold~\cite{wu2021tfrom, fairrecplus, zafar2019fairness, fairrec}. 
\end{enumerate*}
Despite achieving notable performance improvements, existing fairness intervention methods are often designed to optimize specific fairness metrics and typically involve high computational costs, making them challenging to adapt to real-world industrial systems.



\smallskip\noindent%
\textbf{An economic perspective on fair re-ranking.}
In economics, resource allocation typically occurs through processes of distribution and re-distribution~\cite{lambert1992distribution}.  Previous work~\cite{saito2022fair} regards fair ranking as a resource allocation problem and formulated the problem related to Nash Social Welfare in economics, see also~\citet{fairrec, fairrecplus}. TaxRank~\cite{TaxRank} regards fair re-ranking as a taxation process, which often serves as a key mechanism in the re-distribution process, enabling wealth reallocation and addressing income inequality~\cite{hanlon2010review, nerre2001concept}. However, they merely use economic objectives to define different fairness metrics, without understanding how fairness-accuracy trade-offs occur under different metrics.

\section{Problem Formulation}\label{sec:formulation}
We begin by defining the fair re-ranking task, followed by introducing the concept of elasticity in economics. 

\subsection{Fair re-ranking}

In re-ranking tasks, let $\mathcal{U}$ denote the set of users, $\mathcal{I}$ the set of items, and each item $i \in \mathcal{I}$ belongs to a unique group $g \in \mathcal{G}$. The set of items within a specific group $g$ is represented as $\mathcal{I}_g$. When a user $u \in \mathcal{U}$ accesses the re-ranking system, the system will re-rank items more fairly according to a given candidate ranked list of size $K$, denoted as $L_K(u) \in \mathcal{I}^K$. In each ranked list $L_K(u)$, we will get the ranking scores $s_{u,i}, i\in L_K(u)$ generated by ranking models. The ranking score can usually be regarded as the 
probability of a user clicking on an item (\ie click-through-rate (CTR) value~\cite{yang2019bid, liu2021neural}).

Then, we will define the user and item utilities for a certain group in re-ranking tasks. The item group utility $\bm{v}_g$ and user utility $\bm{w}_u$ in re-ranking tasks is typically defined as the accumulated utilities of item group $g$ across all ranked lists: 
\begin{equation}\label{eq:utility_define}
    \bm{v}_g = \sum_{u\in\mathcal{U}}\sum_{i\in  L_K(u)} s_{u,i}I(i\in \mathcal{I}_g), \quad \bm{w}_u = \sum_{i \in L_K(u)} s_{u,i}, 
\end{equation}
where $I(\cdot)$ is the indicator function.




The goal of fair re-ranking $f$ is to maximize the overall user utilities ($\sum_{u \in \mathcal{U}} \bm{w}_u$), while simultaneously striving to equalize utilities across item groups ($\bm{v}_{r} \approx \bm{v}_{p}, \forall i\in\mathcal{I}_{r}, \forall j\in\mathcal{I}_{p}$)

\begin{table}[t]
    \caption{Correspondence between taxation elements in economics and fair re-ranking. }
    \label{tab:compare}
     \small
    \centering
    \begin{tabular}{l l}
    \toprule
    \textbf{Economics} & \textbf{Fair re-ranking}\\
    \hline 
    Consumer (buy product)  & Users $\mathcal{U}$ (click items) \\
    Supplier (sell product) & Item groups $\mathcal{G}$ (provide items) \\
    Commodity tax & Fairness constraint \\
    Tax subsidies for the poor & Increase ranking score for the poor\\
    Selling price (tax objective) & Ranking scores (fairness objective) \\
    Elasticity on price $E_e$ & Elasticity on utilities of item group $E_{r,p}$ \\
    \bottomrule
    \end{tabular}  
    \vspace{-0.4cm}
\end{table}

\subsection{Elasticity theory}
\label{sec:elastic_theory}
In this section, we will first introduce elasticity theory from economics. Then we relate elasticity theory to re-ranking tasks.

\subsubsection{Elasticity in economics.} Elasticity is a measure of the responsiveness of one variable to changes in another variable~\cite{robbins1997elasticity}. 
In economics, the price elasticity of demand is defined as: 
\[
    E_e = \frac{\partial Q/Q}{\partial P/P},
\]
where $Q$ is the user demand quantity of an item and $P$ is the item price. Elastic $E_e$ reflects how sensitive consumers are to changes in price or other variables, like the price of related goods. Specifically, Elastic $E_e$ measures the percentage decrease in the quantity 
$Q$ of items users are willing to purchase when the current price $P$ increases by 1\%.

For example, bread has low elasticity since bread is a necessity for most people. If the price of bread increases slightly, consumers are unlikely to stop buying it. This small change in demand despite a large price change illustrates low elasticity. On the other hand, diamonds are not a necessity and there are many alternatives. If the price of a pair increases slightly, consumers might decide not to buy it and instead look for a cheaper option. This large change in demand with a small price change demonstrates high elasticity. 

This transfer rate of commodity tax burden depends on the price elasticity of demand $E_d$ in taxation theory~\cite{hanlon2010review}. If consumers are less sensitive to price changes, a larger share of the commodity tax burden will fall on consumers because they will continue buying the product even if the price increases due to the tax. On the other hand, if consumers are highly sensitive to price changes, a larger share of the tax burden falls on suppliers because consumers will significantly reduce purchases if prices rise. 

Therefore, when imposing commodity tax, it is generally more effective to tax products with high elasticity, as this minimizes the extent to which the tax burden is transferred to users.

\subsubsection{Elasticity in re-ranking}
When we relate re-ranking to the economic transaction process, we can relate the re-ranking elements to taxation elements in Table~\ref{tab:compare}. Then the utility elasticity  of different two groups (group $r$ and group $p$) of ranking is defined as:
\begin{equation}\label{eq:elastic}
    E_{r,p} = \frac{\partial v_r}{\partial v_p}.
\end{equation}
The elasticity term \( E_{r,p} \) quantifies the sensitivity of item group \( r \)'s utility to changes in item group \( p \)'s utility, capturing the interdependence between the two items in the ranking system.~\footnote{Note that in ranking systems, we focus more on the absolute value rather than the percentage value; thus, the elastic derivation differs slightly from economic elasticity.} 

Intuitively, applying the economics analogy of bread and diamonds to re-ranking, we find a striking similarity: for a relatively rich item group $r_1$ and a less ``rich'' item group $r_2$, they resemble diamonds and bread (with the former having higher elasticity and the latter lower elasticity compared to same group $p$). Suppose we slightly reduce the exposure of $r_1$ to subsidize the poor item group $p$, then $p$'s utility would increase significantly with minimal accuracy loss. 
In contrast, reducing some exposure from $r_2$ to subsidize $p$ would lead to a smaller increase in $p$'s utility but would transfer most of the ``tax'' to users, resulting in more accuracy loss. 


Therefore, in fair re-ranking, we should reduce more exposure from item groups with higher elasticity to subsidize the poorer groups. Detailed theoretical analysis can be seen in the next section.

\section{Fairness Evaluation in Re-ranking}\label{sec:evaulation}



We first analyze the fairness objectives of fair re-ranking using elasticity theory. Then, we propose a new fairness evaluation metric EF by introducing the EF-Curve.

\subsection{Analysis of fairness metrics}
\label{sec:fair_objective}
We aim to demonstrate that different fairness objectives fundamentally alter the elasticity of different item groups, reflecting their varying adaptability to fairness constraints.

Firstly, we define the fairness objective. Let $\bm{v} = [\bm{v}_1, \bm{v}_2, \ldots, \bm{v}_{|\mathcal{G}|}]$ be the utility vector of different item groups. The fairness objective involves defining a function $f(\bm{v})$, where the output represents a fairness metric that quantifies the inequity among item group utilities. The function $f(\bm{v})$ should increase when different utilities become more equal. $f(\bm{v})$ usually has many different forms in previous work, such as max-min fairness~\cite{xu2023p}, entropy fairness~\cite{deldjoo2019recommender}, $\alpha$-fairness~\cite{TaxRank}, proportional fairness~\cite{wei2022rank}, p-norm~\cite{bektacs2020using} and Renyi Entropy~\cite{renner2004smooth}. Previous studies often adopted different fairness optimization objectives and evaluation metrics without fully understanding the distinctions between these metrics.

\subsubsection{General form of fairness metric}

From the perspective of taxation~\cite{TaxRank}, fairness in re-ranking objectives imposes taxes on richer item groups and redistributes these as subsidies to poorer item groups. Given the requirements of the taxation~\cite{nerre2001concept}, the fairness function $f(\bm{v})$ should be: (i) continuous on $\mathbb{R}_{+}$; (ii) scale-invariant: $f(\bm{v}) = f(c\bm{v}), \forall c>0$; and (iii) independent of number of item groups. 
Also, given the requirements of ranking systems, the fairness optimization objective should be distributed~\cite{fairsync}, allowing the data to be partitioned across different servers and aggregated results to ensure scalability and efficiency.

\begin{figure}[t]  
    \centering    
    \includegraphics[width=0.95\linewidth]{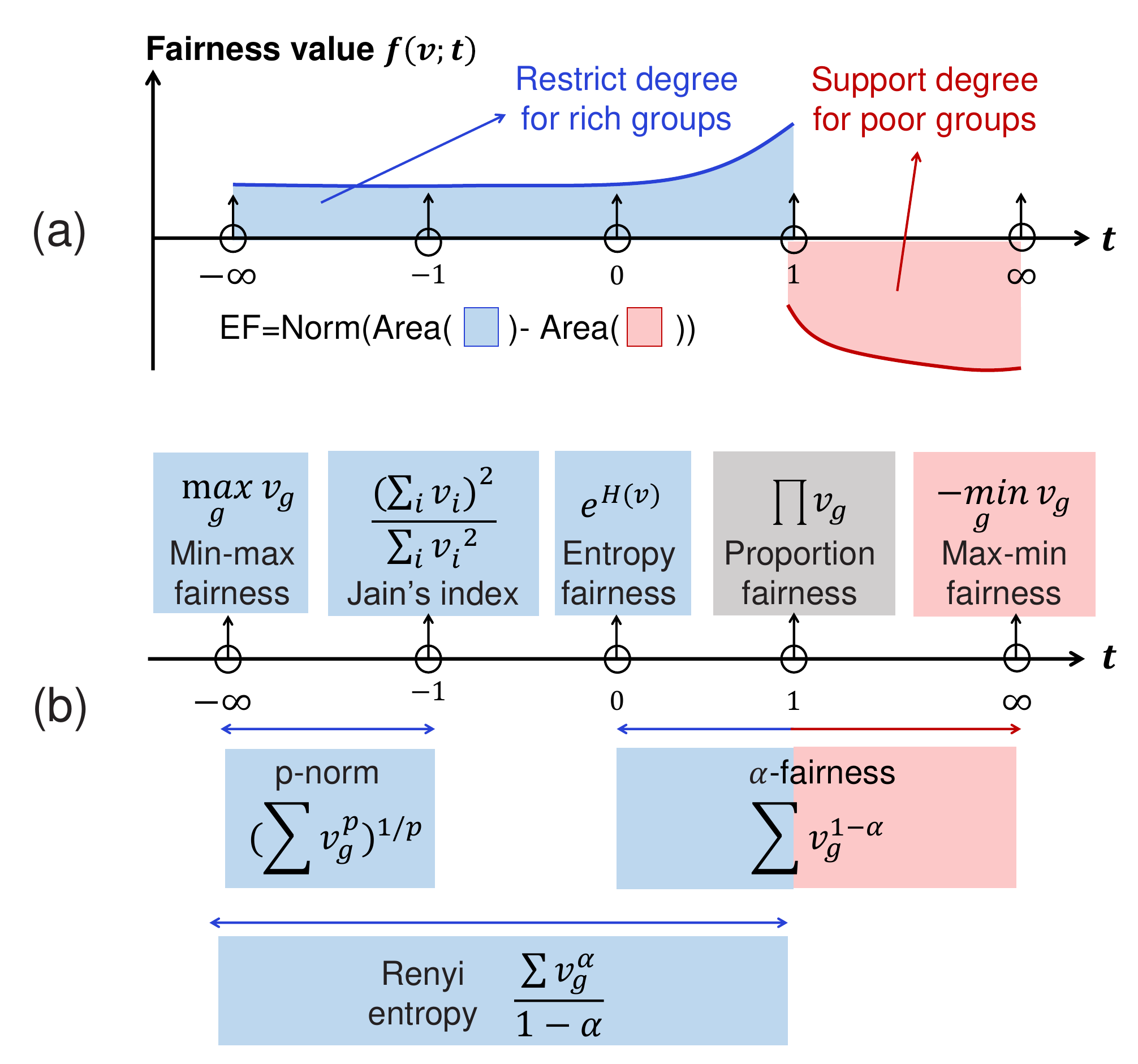}
    \caption{(a)~The EF-Curve, where the x-axis is tax base $t$ and the y-axis is the fairness metric $f(\bm{v};t)$. EF-Curve describes the restrict/support degree for different item groups. (b)~The specific fairness metrics corresponding to particular $t$ values. }
    \label{fig:EF_Curve}
    \vspace{-0.3cm}
\end{figure}

According to 
these requirements, let $\bar{\bm{v}}_g$ be the normalized utility: $\bar{\bm{v}}_g=\bm{v}_g/\sum_{g=1}^{|\mathcal{G}|} \bm{v}_g$,  then fairness metrics have a general form:
\begin{equation}\label{eq:general_form}
    f(\bm{v}; t) = \text{sign}(1-t)\left(\sum_{g=1}^{|\mathcal{G}|} \bar{\bm{v}}_g^{1-t}\right)^{\frac{1}{t}},
\end{equation}
where $\text{sign}(\cdot)$ is the symbolic function and $t$ is the tax base, illustrated in the following theorem:


\begin{theorem}
\label{theo:general_fair}
    The $f(\bm{v}; t)$ is the unique form of $f(\bm{v})$. When $t$ takes on different values, as shown in Figure~\ref{fig:EF_Curve}(b), $f(\bm{v};t)$ will be generalized to different fairness metrics, especially,
    $
    \lim_{t \to 0}f(\bm{v}; t) = e^{H(\bar{\bm{v}})},
    $
    where $H(\bar{\bm{v}})$ is the entropy fairness:
    $
        H(\bar{\bm{v}}) = - \sum_{g=1}^{|\mathcal{G}|}\bar{\bm{v}}_g\log \bar{\bm{v}}_g.
    $
\end{theorem}

\noindent%
A detailed proof of Theorem~\ref{theo:general_fair} is provided in Appendix~\ref{app:proof_general_fair}. Eq.~(\ref{eq:general_form}) presents a general form of fairness metric, where the parameter $t$ can be adapted to represent various fairness metrics, as illustrated in Figure~\ref{fig:EF_Curve}(b).
The parameter $t$ can be understood as the tax base, which will be further interpreted through the elastic theory described in the next section. Intuitively, as the absolute value $|t|$ increases, the fairness metric places greater emphasis on item groups with lower utility values.

\subsubsection{Elasticity theory for analyzing the general form.} 
In this section, we apply the elasticity theory introduced in Section~\ref{sec:elastic_theory} to analyze the general form of the fairness metric described in Eq.~(\ref{eq:general_form}) through the following theorem:

\begin{theorem}\label{theo:tax_t}
    The parameter $t$ in Eq.~(\ref{eq:general_form}) represents the tax base, where, upon adding the next item to the ranked list, the rich group $r$  compared to the poor group $p$ will be subjected to a commodity tax with a rate of:
    \begin{equation}
    E_{r,p} = \frac{\partial v_p}{\partial v_r} = \left(\frac{\bar{\bm{v}}_r}{\bar{\bm{v}}_p}\right)^{-|t|},
    \end{equation}
    which means the elastic value as defined in Eq.~(\ref{eq:elastic}). Meanwhile, the rich and poor item group threshold is:
   $
        \theta = \left(\frac{\sum_{g=1}^{|\mathcal{G}|}v_g}{\sum_{g=1}^{|\mathcal{G}|}v_g^{1-t}}\right)^{\frac{1}{t}},
    $
    where $\bm{v}_g>\theta$, the group $g$ is a rich group, otherwise, it will be a poor group.
\end{theorem}

\noindent%
A detailed proof can be found in Appendix~\ref{app:proof_tax_t}. 
Theorem~\ref{theo:tax_t} means that different fairness metrics change the elasticity between the poor item group and the rich item group. 

Intuitively, different fairness metrics in re-ranking tasks are determined by adjusting the elasticity between groups. From a taxation perspective, these metrics impose $E_{r,p}$ times more commodity tax to the rich groups $r$ and subsidy to the poor group $p$.

\subsection{EF-Curve and EF metric}
After analyzing fairness metrics from an economic perspective, we observe that previous fair re-ranking evaluations rely on single fairness metrics (a single 
$t$ value), which assesses tax transfers only under specific elasticity conditions.

To provide a comprehensive fair re-ranking evaluation, we use the general form $f(\bm{v};t)$ in Eq.~(\ref{eq:general_form}) to design the EF-curve (shown in Figure~\ref{fig:EF_Curve} (a)). On the EF-curve, each point on the horizontal axis corresponds to a different tax base $t$, and each point on the vertical axis reflects the fairness metric $f(\bm{v};t)$ under the respective metric. 

Comparing the EF-Curve of different algorithms reveals their performance across varying elasticities, helping to identify the most suitable algorithm for online deployment based on specific application requirements. Meanwhile, we propose to utilize the area enclosed by the EF-curve (shown in Figure~\ref{fig:EF_Curve} (a)) and the axes (called EF):
\begin{equation}~\label{eq:EF}
    \text{EF} = \int_{1-M}^{1+M} \frac{f(\bm{v};t)}{Z} dt,
\end{equation}
where $Z=2M|\mathcal{G}|$ is the normalized factor for the area computation (\ie the Norm operation in Figure~\ref{fig:EF_Curve} (a)), $M\geq 0$ is used to approximate infinity integral values.
Intuitively, Eq.~(\ref{eq:EF}) measures the averaged fairness performances across different fairness metrics.

\section{Accuracy-Fairness Optimization}\label{sec:method}

We first analyze the accuracy-fairness ranking trade-off objective from an economic perspective. Then we propose a new fair re-ranking algorithm named ElasticRank.

\subsection{Analysis of the accuracy-fairness trade-off}
\label{sec:trade-off}

Given the general form of the fairness metric in Theorem~\ref{theo:general_fair}, this section aims to analyze the objective of maximizing ranking accuracy while balancing the trade-off with the fairness function. We aim to show that the trade-off is just like the commodity tax transfer process in economics, using elasticity theory.

In ranking tasks~\cite{xu2023p, cpfair}, previous work often adapts a linear trade-off between the fairness and accuracy function:
\begin{equation}\label{eq:tradeoff_v1}
    v^{*} = \argmax_{\bm{v}} \sum_{u\in\mathcal{U}}\bm{w}_u + \lambda f(\bm{v}),
\end{equation}
where the $\bm{w}_u$ is defined as Eq.~(\ref{eq:utility_define}) and $\lambda\in [0,\infty)$ is the trade-off co-efficient and the first part is the accuracy part while the second part is the fairness objective. Next, we will use the following theorem to rewrite and analyze the trade-off function.

\begin{theorem}\label{theo:trade-off}
    Eq.~(\ref{eq:tradeoff_v1}) can equivalently be optimized as:
    \begin{equation}
        v^{*} = \argmax_{\bm{v}} L = \argmax_{\bm{v}} f(\bm{v};|t|)^{|t|} \cdot a(\bm{w})^{1-|t|},
    \end{equation}
    where $a(\bm{w}) = \sum_{u\in\mathcal{U}} \bm{w}_u$ is the accuracy function. 
    
    Let $\eta=\bm{1}-\bar{\bm{v}}$ be the gradient direction of nearest fairness (moving to the averaged utility) and $\alpha = \nabla_{\bm{v}} \bm{w}$ be the gradient of accuracy function. The transfer ratio between fairness (commodity tax on groups) and accuracy (commodity tax transferred to users) is:
    \begin{equation}\label{eq:tradeoff_radio}
        \gamma = \frac{\langle \nabla_{\bm{v}} L,\eta \rangle}{\langle \nabla_{\bm{v}} L,\alpha \rangle} =1-\frac{1}{1+k(E_{r,p})},  ~ k(E_{r,p}) = \frac{\sum_{p\in\mathcal{G}}\sum_{r\neq p}\bm{v}_p^{1-|t|}E_{r,p}}{\sum_{p\in\mathcal{G}} \bm{v}_p^{1-|t|}}.
    \end{equation} 
    The transfer ratio $\gamma$ is determined by the elasticity between any two item groups, where the ratio can also be interpreted as the extent to which the commodity tax is transferred to the users.
\end{theorem}

\noindent%
A detailed proof is provided in Appendix~\ref{app:proof_trade-off}. 

In Figure~\ref{fig:tradeoff} (a), we give a more intuitive example to understand how elasticity works for the commodity tax transfer. For example, Jain's index can naturally be regarded as changing the elasticity $E_{r,p}=1$ (w/o fairness) of two different item groups as $\left(\frac{\bar{\bm{v}}_r}{\bar{\bm{v}}_p}\right)^{-1}$. Suppose there are two item groups, 
$p$ with a utility of $1$ and $r$ with a utility of $3$. In that case, Jain's index indicates that adding the next item from group $p$ to the ranked list, compared to group $r$, will have its utility weighted three times more to support the poorer group (changing elasticity $E_{r,p}=3$). 

From a taxation perspective, as shown in Figure~\ref{fig:tradeoff} (b), Jain's index will give three times more commodity tax to the rich groups than to the poor group. However, the commodity tax (item fairness) will inevitably be transferred to accuracy loss. Intuitively, when $E_{r,p}$ is small, poor items act like necessities, adding exposure yields limited systemic utility and maximizes accuracy loss (lower commodity tax transfer rate). When $E_{r,p}$ is large, poor items resemble luxuries, adding exposure greatly boosts systemic utility but increases accuracy loss (higher commodity tax transfer rate). 


\begin{figure}[t]  
    \centering    
    \includegraphics[width=\linewidth]{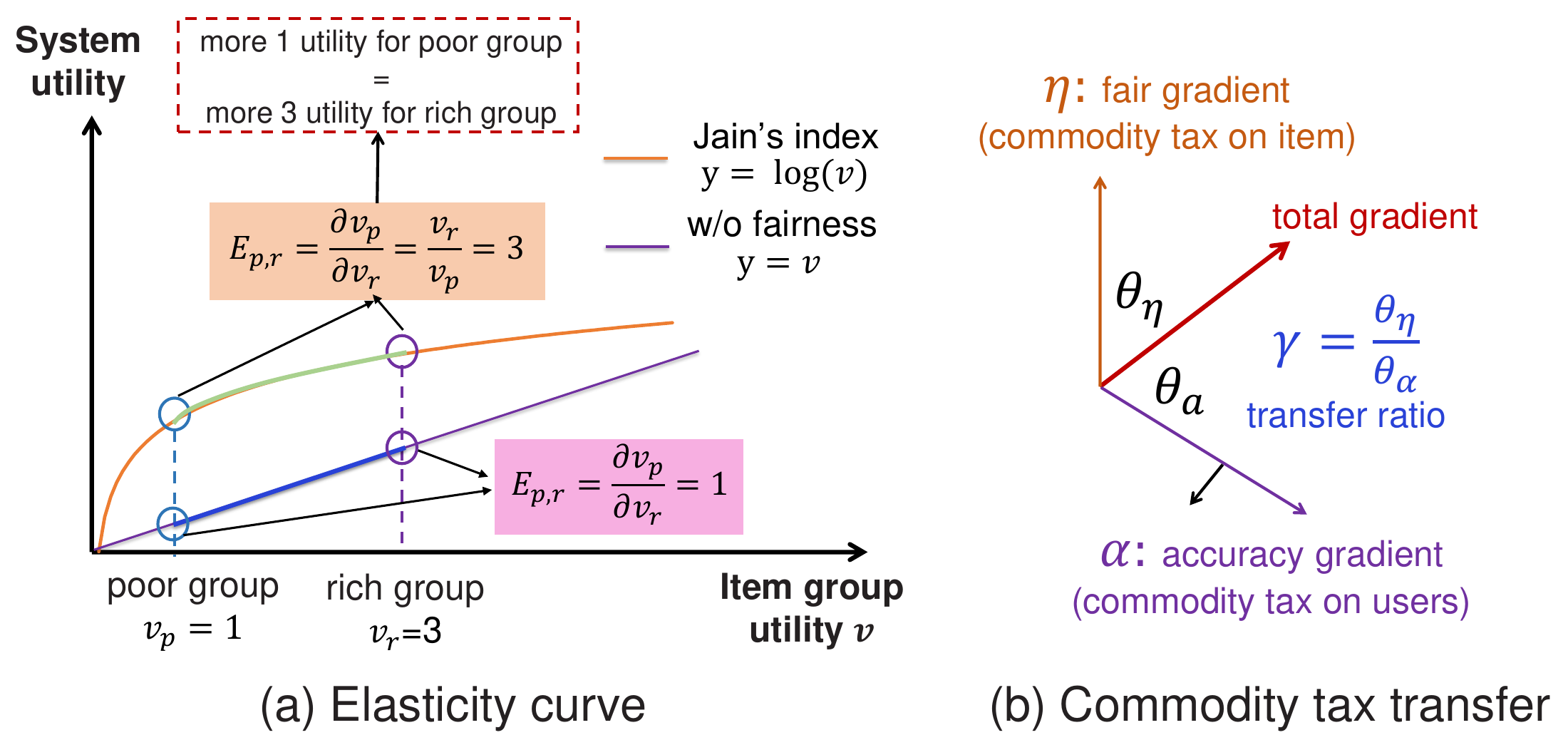}
    \caption{(a)~The Elasticity curve when optimizing Jain's index and objective without fairness constraint. (b)~Illustration of how the commodity tax (item fairness) is being transferred to users (accuracy loss).}
    \label{fig:tradeoff}
\end{figure}

\subsection{ElasticRank}
Inspired by the aforementioned analysis, we propose an efficient and effective re-ranking algorithm, ElasticRank. 

From the proof of Theorem~\ref{theo:trade-off}, we observe that the fairness function influences the optimal gradient direction, steering it to balance the trade-off between fairness and accuracy. Moreover, we can observe that as the elasticity value increases, the ratio of commodity tax transfer also rises, reflecting a stronger responsiveness of the tax structure to changes in elasticity. 
From an optimization perspective, the elasticity in economics essentially curves the optimization space, increasing the separation between two item groups, as represented by their geodesic distance~\cite{bouttier2003geodesic}. Therefore, we propose ElasticRank, which measures item distances in the elasticity-curved space.

The overall workflow can be seen in Algorithm~\ref{alg:elasticrank}. Formally, the ElasticRank re-ranked list for each user $u$ can be defined as
\begin{equation}\label{eq:ranking_result}
    L_K(u) = \argmax_{i\in\mathcal{I}^k} \sum_{i\in\mathcal{I}} \left[s_{u,i}+d\left(g(i),a\right)\right],
\end{equation}
where $g(i)$ is the group of item $i$ and $d(g(i),a)$ represents the additional distance in the curved space between the group $g(i)$ and an anchor group $a$. The anchor group $a$ can be any group in order to reduce the computational complexity. 
In our algorithm, $a$ is chosen as the group that has the last $\eta \%$ utility when the user $u$ arrives in the systems:
$
    a = \text{argsort}(\bm{v})[b], ~b=\eta\%|\mathcal{G}|.
$

The additional distance $ d\left(g(i),a\right)$ is computed through the curve distance:
\begin{equation}
\label{eq:curve_distance}
\mbox{}
\hspace*{-1mm}
    d\left(g(i),a\right) = \!\int_{v_{g(i)}}^{v_{a}} \!\sqrt{1+\left(\frac{\partial \bm{v}_a}{\partial x}\right)^2}dx \approx (1-t)\bm{v}_{g(i)}^{-t}(\bm{v}_a^{1-t}-\bm{v}_{g(i)}^{1-t}),
\end{equation}
where the term $\frac{\partial \bm{v}_a}{\partial x}$ measures the variation in the slope of the curve (i.e., the elasticity $E_{a,g(i)}$ of group $a$ with respect to group $g(i)$ moves along the curve). Intuitively, if the utility of group $g(i)$ has a huge gap with the anchor group $a$, then the distance $d\left(g(i),a\right)$ will be larger to close the utility of such two groups. 

\begin{algorithm}[t]
    \caption{Learning algorithm of ElasticRank}
	\label{alg:elasticrank}
	\begin{algorithmic}[1]
	\REQUIRE User set $\mathcal{U}$, item set $\mathcal{I}$, group set $\mathcal{G}$, ranking size $K$, tax rate $t$, anchor group index $b=\eta\% |\mathcal{G}|$, user-item ranking score $s_{u,i}, \forall u\in\mathcal{U},\forall i\in\mathcal{I}$ 
        \STATE Set $\bm{v}_g = 1, \forall g\in\mathcal{G}$
        \FOR{$u\in\mathcal{U}$}
            \STATE Choose anchor group $a=\text{argsort}(\bm{v}){[b]}$
            \STATE $d\left(g(i),a\right) = (1-t)\bm{v}_{g(i)}^{-t}(\bm{v}_a^{1-t}-\bm{v}_{g(i)}^{1-t})$
            \STATE $L_K(u) = \argmax_{i\in\mathcal{I}^k} \sum_{i\in\mathcal{I}} \left[s_{u,i}+d\left(g(i),a\right)\right]$
            \STATE $\bm{v}_g = \bm{v}_g + \sum_{i\in\mathcal{I}_g} s_{u,i}, \forall g\in\mathcal{G}$
        \ENDFOR
	\end{algorithmic}
\end{algorithm}

Intuitively, taxing higher elasticity item groups minimizes accuracy loss. ElasticRank leverages elasticity calculations to adjust item distances in a curved space, boosting fairness scores for high-elasticity groups and reducing accuracy loss.
Meanwhile, as shown in Algorithm~\ref{alg:elasticrank}, ElasticRank does not introduce additional computational overhead, ensuring that its complexity aligns with that of standard ranking algorithms.

\section{Experiments}
We evaluate ElasticRank using three publicly available ranking datasets, and the source code is shared at GitHub~\url{https://github.com/XuChen0427/ElasticRank}.

\begin{table*}[ht]
\setlength{\tabcolsep}{4pt}
        \Small
        \caption{Performance comparison between ElasticRank and the baselines on Steam, Amazon, and Yelp, where we tuned various re-ranking models to achieve accuracy performances (NDCG) close to 99\%, and evaluated fairness across three cut-offs $K$ to assess effectiveness. Bold numbers mean the best fairness performance for the models controlling NDCG close to 99\% (except for FairRec, FairRec+, and TaxRank); $\uparrow$ and $\downarrow$ indicate whether a higher or lower value of the metric is better, respectively. $*$ indicates that the improvements over the baselines are statistically significant (t-tests and $p$-value $< 0.05$).}
    \label{tab:EXP:main}
    \centering
\begin{tabular}{@{} l  l lll lll lll @{}}
\toprule
\multicolumn{1}{c}{} & \textbf{\text{Top-K}} & \multicolumn{3}{c}{\textbf{K=5}} & \multicolumn{3}{c}{\textbf{K=10}} & \multicolumn{3}{c}{\textbf{K=20}} \\ 
\cmidrule(r){3-5}
\cmidrule(r){6-8}
\cmidrule{9-11}
\multicolumn{1}{c}{} & \textbf{Models} & \multicolumn{1}{c}{Loss@5$\downarrow$} & \multicolumn{1}{c}{NDCG@5$\uparrow$} & EF@5$\uparrow$ & \multicolumn{1}{c}{Loss@10$\downarrow$} & NDCG@10$\uparrow$ & EF@10$\uparrow$ & \multicolumn{1}{c}{Loss@20$\downarrow$} & \multicolumn{1}{c}{NDCG@20$\uparrow$} & EF@20$\uparrow$ \\ 
\midrule
\multirow{8}{*}{\textbf{Steam}} & FairRec & 0.1724 & 0.9395 & -2.0529 & 0.0833 & 0.9670 & -1.9659 & 0.0740 & 0.9653 & -2.2483 \\
 & FairRec+ & 0.1500 & 0.9470 & -2.1579 & 0.0836 & 0.9668 & -1.9799 & 0.0785 & 0.9632 & -2.2416 \\
 & TaxRank & 0.1105 & 0.9468 & -1.2284 & 0.1183 & 0.9374 & -1.0655 & 0.1236 & 0.9249 & -0.7544 \\ 
 \cmidrule{2-11}
 & Welf & 0.0491 & 0.9760 & -1.1725 & 0.0409 & 0.9776 & -1.0188 & 0.0283 & 0.9812 & -0.8818 \\ 
 & CPFair & 0.0251 & 0.9891 & -1.3717 & 0.0174 & 0.9911 & -1.5670 & 0.0109 & 0.9938 & -1.3234 \\ 
 & P-MMF & 0.0286 & 0.9875 & -1.0435 & 0.0222 & 0.9890 & -0.9228 & 0.0221 & 0.9879 & -0.7744 \\ 
 & min-regularizer & 0.0219 & 0.9903 & -1.4547 & 0.0368 & 0.9797 & -1.3490 & 0.0374 & 0.9747 & -0.9249 \\ 
 & \textbf{ElasticRank (Ours)} & 0.0174 & 0.9924 & \textbf{-0.9147}$^*$ & 0.0112 & 0.9948 & \textbf{-0.8283}$^*$ & 0.0136 & 0.9931 & \textbf{-0.7310}$^*$ \\ 
\midrule
\multirow{8}{*}{\textbf{Amazon}} & FairRec & 0.0000 & 1.0000 & -26.5456 & 0.0000 & 1.0000 & -12.9110 & 0.0012 & 0.9990 & -7.7857 \\ 
 & FairRec+ & 0.0000 & 1.0000 & -26.5456 & 0.0000 & 1.0000 & -12.9110 & 0.0013 & 0.9990 & -8.1340 \\ 
 & TaxRank & 0.2827 & 0.7098 & -0.8586 & 0.1526 & 0.7992 & -0.9156 & 0.0508 & 0.9150 & -1.1968 \\ 
  \cmidrule{2-11}
 & Welf & 0.0315 & 0.9734 & -3.6578 & 0.0326 & 0.9679 & -0.9395 & 0.0096 & 0.9902 & -0.8497 \\ 
 & CPFair & 0.0112 & 0.9910 & -2.7632 & 0.0064 & 0.9942 & -1.1135 & 0.0068 & 0.9929 & -0.8907 \\ 
 & P-MMF & 0.0107 & 0.9916 & -2.4519 & 0.0092 & 0.9917 & -1.0574 & 0.0091 & 0.9907 & -0.8741 \\
 & min-regularizer & 0.0108 & 0.9906 & -2.9959 & 0.0380 & 0.9543 & -1.0571 & 0.0147 & 0.9813 & -0.8615 \\ 
 & \textbf{ElasticRank (Ours)} & 0.0102 & 0.9905 & \textbf{-2.4391}$^*$ & 0.0061 & 0.9940 & \textbf{-1.0370}$^*$ & 0.0095 & 0.9898 & \textbf{-0.8334}$^*$ \\ 
 \midrule
\multirow{8}{*}{\textbf{Yelp}} & FairRec & 0.0413 & 0.9715 & -7.5726 & 0.0176 & 0.9883 & -7.6010 & 0.0145 & 0.9905 & -7.5270 \\
 & FairRec+ & 0.0064 & 0.9962 & -14.2024 & 0.0048 & 0.9972 & -11.4978 & 0.0047 & 0.9971 & -10.0469 \\
 & TaxRank & 0.0665 & 0.9331 & -0.5671 & 0.0363 & 0.9588 & -0.6460 & 0.0319 & 0.9657 & -0.7348 \\ 
  \cmidrule{2-11}
 & Welf & 0.0147 & 0.9870 & -0.5316 & 0.0086 & 0.9932 & -0.4623 & 0.0123 & 0.9906 & -0.4011 \\
 & CPFair & 0.0173 & 0.9885 & -0.5281 & 0.0091 & 0.9927 & -0.4584 & 0.0128 & 0.9901 & -0.3951 \\ 
 & P-MMF & 0.0085 & 0.9933 & -0.5302 & 0.0098 & 0.9923 & -0.4379 & 0.0159 & 0.9876 & -0.3350 \\ 
 & min-regularizer & 0.0122 & 0.9888 & -0.5544 & 0.0110 & 0.9900 & -0.4438 & 0.0158 & 0.9876 & -0.3357 \\
 & \textbf{ElasticRank (Ours)} & 0.0116 & 0.9908 & \textbf{-0.5257}$^*$ & 0.0108 & 0.9918 & \textbf{-0.4211}$^*$ & 0.0154 & 0.9901 & \textbf{-0.3046}$^*$ \\ 
 \bottomrule
\end{tabular}
    
\end{table*}

\subsection{Experimental settings}

\textbf{Dataset.} Our experiments are based on three large-scale, publicly available ranking applications, including:

$\bullet$
\textbf{Steam}~\cite{SASRec}: a ranking dataset for games on the Steam platform.
We use the data for games played for more than 10 hours in our experiments. The publishers of games are considered item groups. It has 169,030 samples, which contains 4,446 users, 1,238 items, and 43 item groups.\footnote{\url{http://cseweb.ucsd.edu/~wckang/Steam_games.json.gz}.}

$\bullet$
\textbf{Amazon-Digital-Music}~\citep{he2016ups}: a subset (digital music domains) of Amazon Product dataset. After the pre-processing steps, it has 11,320 samples, with 3,175 users, 3,766 items, and 26 item groups.\footnote{\url{http://jmcauley.ucsd.edu/data/amazon/}.}

$\bullet$ \textbf{Yelp}: a large-scale businesses recommendation dataset. The categories of items are considered as item groups.  After the pre-processing steps, it has 702,457 samples, which contains 8,198 users, 6,429 items, and 64 item groups.\footnote{\url{https://www.yelp.com/dataset}.}

During the pre-processing step, users and items that have interactions with fewer than $L$ items or users are excluded from the entire dataset to mitigate the issue of extreme sparsity. For Yelp, we set $L=10$ and for the other two datasets, we set $L=5$. Following~\citet{fairsync}, we consider groups with fewer than 10 items as a single group, which we name the ``infrequent group''.

Following~\cite{xu2023p, cpfair}, we sort all interactions by time and use the first 80\% of the interactions as data to train the base ranking model (i.e., MF ranking model~\cite{DMF}). The remaining 20\% of interactions are used as the test data for re-ranking tasks.

\textbf{Evaluation.} The performance of the models is evaluated from two aspects: re-ranking accuracy and fairness degree. 
For the accuracy, 
following~\cite{xu2023p, wu2021tfrom}, we use NDCG@K and Loss@K:
\begin{align}
   \text{NDCG@K} & =\frac{1}{|\mathcal{U}|} \sum_{u\in\mathcal{U}}\frac{\sum_{i\in\mathbf{L}_K^F(u)}s_{u,i}/\log(\textrm{rank}^F_i+1)}{\sum_{i\in\mathbf{L}_K(u)}s_{u,i}/\log(\textrm{rank}_i+1)},
\\
     \text{Loss@K} & = \frac{\sum_{u\in\mathcal{U}}\sum_{i\in\mathbf{L}_K(u)}s_{u,i}-\sum_{u\in\mathcal{U}}\sum_{i\in\mathbf{L}_K^F(u)}s_{u,i}}{|\mathcal{U}|K},
\end{align}
where $\mathbf{L}_K(u_t)$ is the original ranked list and $\mathbf{L}^F_K(u_t)$ is the fair-aware re-ranked list, and rank$_i$ and rank$_i^F$ are the ranking positions of the item $i$ in $\mathbf{L}_K(u_t)$ and $\mathbf{L}_K^F(u_t)$, respectively.  The improved re-ranking accuracy results in a higher NDCG@K value and a lower Loss@K value, indicating better re-ranking quality.


For the fairness degree, we use the EF metric defined in Eq.~(\ref{eq:EF}). The improved fairness results in a higher EF@K value.

\textbf{Baselines.}
The following representative item fairness re-ranking models were chosen as baselines: \textbf{FairRec}~\cite{fairrec} and \textbf{FairRec+} \cite{fairrecplus} propose to ensure Max-Min Share of exposure for different items. We also compare \textbf{Welf}~\cite{nips21welf}, which uses the Frank-Wolfe algorithm to optimize two-sided fairness. \textbf{CPFair}~\cite{cpfair} formulates the re-ranking problem as a knapsack problem and solves it greedily. \textbf{min-regularizer}~\cite{xu2023p}: adds a regularizer that penalizes the exposure gaps between the target provider and the groups that have worst-off utilities. \textbf{P-MMF}~\cite{xu2023p} uses the mirror gradient descent method to improve the worst-off item group's utility. \textbf{Tax-Rank}~\cite{TaxRank}\footnote{Note that the evaluation of \textbf{Tax-Rank} employs a probability sampling method for re-ranking, whereas we only adopt one sample operation for fairly comparison.} solves the re-ranking utilizing the optimal transportation (OT) techniques.

\textbf{Implementation details.} Our experiments are implemented using Python 3.9. All experiments are conducted on a server with Ubuntu 18.04. As for the hyper-parameters in all models, the anchor group radio $\eta\%$ is tuned among $[50\%,95\%]$. The tax base $t$ is tuned among $[1, 2]$. The range of the integral bounds, denoted as $M$, is set to 50. The ranking base model is the most commonly used MF~\cite{DMF}. 

\subsection{Experimental results}
We report on the performance of ElasticRank and other baselines. 

\subsubsection{Fairness performance comparison.}
To enable fair comparisons, we conduct experiments to show the performance of ElasticRank and other baselines by tuning the accuracy (NDCG@K) close to $99\%$ to test the accuracy-fairness trade-off performances in re-ranking tasks. Since an accuracy loss within 1\% can be considered acceptable for ranking models, as the platform's profit fluctuation remains within a normal range~\cite{BankFair}. Testing fairness improvements within this range is more meaningful and instructive for real-world applications.
Table~\ref{tab:EXP:main} presents the experimental outcomes for our ElasticRank model and the baseline methods across all datasets and all ranking sizes $K=5,10,20$.

From Table~\ref{tab:EXP:main}, we first observe that \textbf{FairRec}~\cite{fairrec}, \textbf{FairRec+}~\cite{fairrecplus}, and \textbf{TaxRank}~\cite{TaxRank}, as item-level fairness methods, struggle to effectively balance accuracy and group-level fairness, often resulting in either insufficient accuracy or inadequate fairness. For the remaining baselines, the experimental results clearly demonstrate that ElasticRank achieves superior fairness performance at the same accuracy level, highlighting the effectiveness of our model.

Next, we test the performance of ElasticRank and the baselines for different accuracy-fairness trade-off degrees and various fairness metrics by conducting experiments on the Steam dataset. Similar trends can be observed in the other two datasets.

\subsubsection{Performance on different tax base $t$.} 
\begin{figure}[!h]  
    \centering    
    \includegraphics[width=\linewidth]{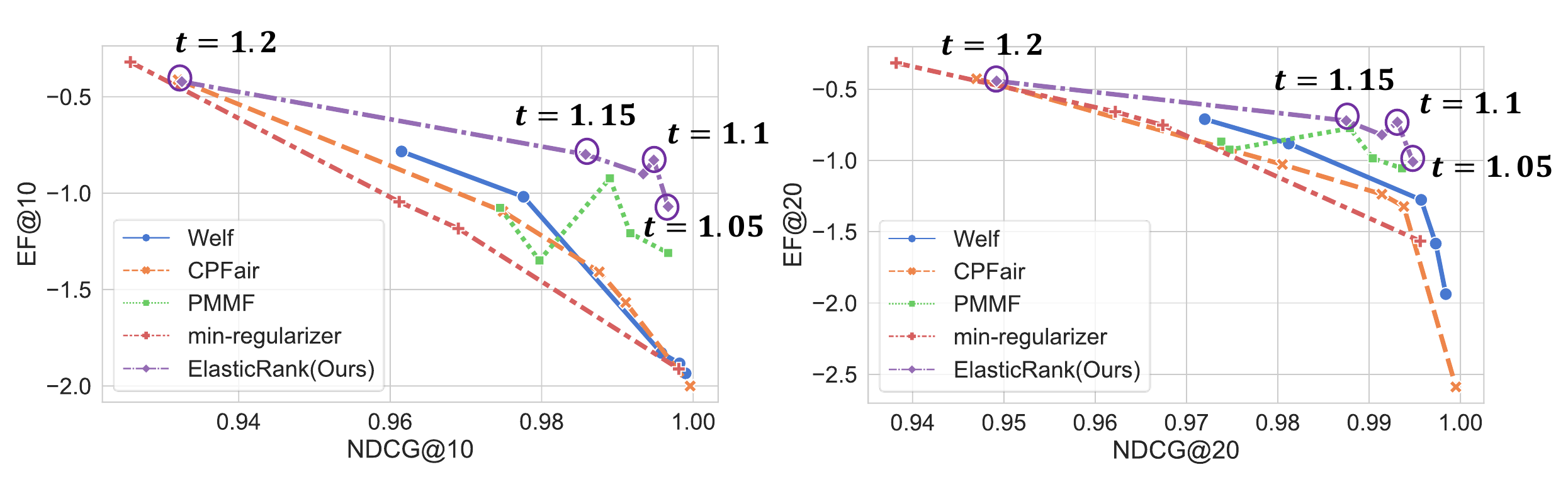}
    \caption{Pareto frontier with different size $K$ under Steam.  }
    \label{fig:Pareto}  
    \vspace{-0.3cm}
\end{figure}

Figure~\ref{fig:Pareto} presents the Pareto frontiers~\cite{TaxRank} for the accuracy metric (NDCG@K) and fairness metric (EF@K) at $K=10$ and $K=20$. These Pareto frontiers~\cite{lotov2008visualizing} are derived by systematically adjusting different model parameters and selecting the points that optimize both NDCG@K and EF@K, thus achieving an ideal trade-off between item fairness and total utility. Note that we only compare the best trade-off baselines, excluding FairRec, FairRec+, and TaxRank.

Firstly, it is evident that a trade-off exists between re-ranking accuracy metrics (NDCG@K) and item fairness metrics (EF@K) with respect to the tax base $t$. When the tax base $t$ is small, FairTax prioritizes ranking accuracy (where item distances are computed in the flat space). Still, when the tax base $t$ increases, ElasticRank emphasizes item fairness by boosting fairness scores for high-elasticity groups while minimizing accuracy loss.

Moreover, compared to the baseline methods, it is evident that the proposed ElasticRank method consistently outperforms them, as indicated by the ElasticRank curves occupying the upper right corner of the Pareto front. This Pareto dominance demonstrates that, for a given NDCG@K level, ElasticRank achieves superior EF@K values, and for a given EF@K level, it delivers better NDCG@K performance. These results highlight the significant advantage of ElasticRank over the baseline methods.

\subsubsection{Performances on EF-Curve}
\begin{figure}[!h]  
    \centering    
    \includegraphics[width=\linewidth]{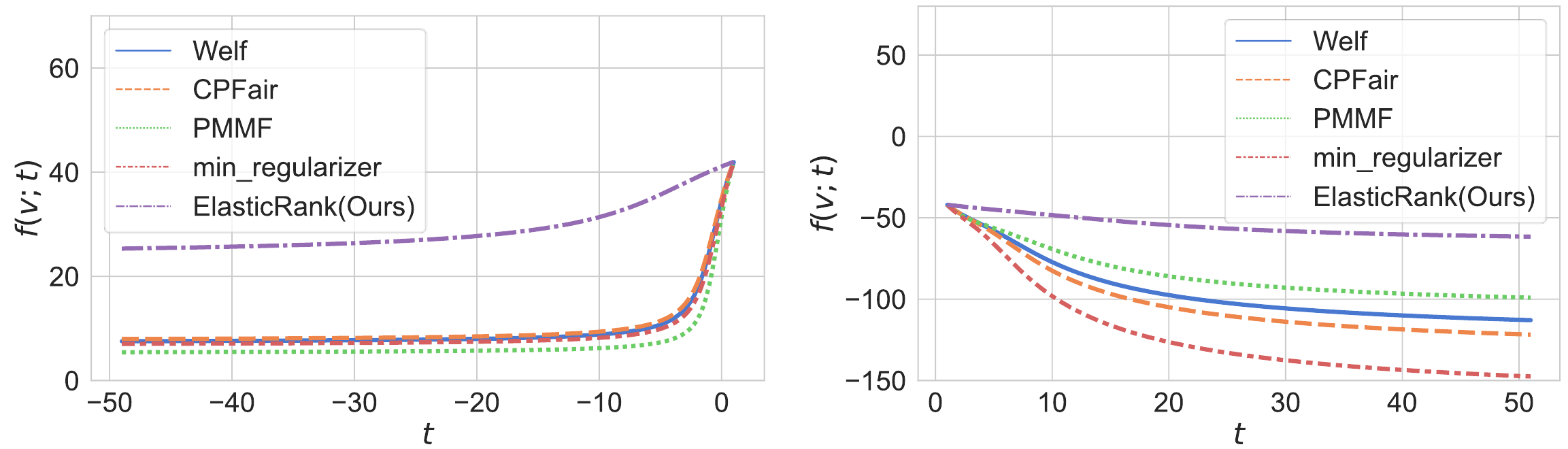}
    \caption{EF-Curve for different models with cut-off size $K=10$ under Steam.}
    \label{fig:exp_EF_Curve}  
\end{figure}

Figure~\ref{fig:exp_EF_Curve} presents the EF-Curve described in Section~\ref{sec:evaulation} on Steam under $K=10$. The experiments were also conducted for controlling NDCG close to 99\% (see Table~\ref{tab:EXP:main}).
Intuitively, each point on the EF-Curve represents a fairness metric that evaluates the level of fairness, with these metrics reflecting the varying degrees of support provided to different item groups, each characterized by different elasticity. When $t$ approaches $-\infty$, the function $f(\bm{v};t)$ measures the utility of the richest groups while when $t$ approaches $+\infty$, the function $f(\bm{v};t)$ measures the utility of the poorest groups. 

Firstly, we observe that different fair re-ranking algorithms enhance fairness from distinct perspectives. For instance, P-MMF often aims to support the poor groups (as indicated by the highest EF-curve of baselines when $t<0$), while struggling to restrict the utility of the rich groups (as shown by the lowest EF-curve when $t<0$). Similarly, we can observe that CPFair is good at restricting the utility of the rich groups but fails to support the poor groups.
Through the EF-Curve, we can observe how each algorithm impacts fairness, allowing us to select the most suitable algorithm based on the specific requirements of different applications.

Moreover, compared to the baseline methods, it is clear that ElasticRank consistently outperforms them at every point on the EF-curve. These results demonstrate that ElasticRank not only effectively restricts the utility of the rich groups (as shown by the highest curve when $t>0$), but also better supports the poor groups (indicated by the highest EF-curve when $t>0$). These findings highlight the significant advantage of ElasticRank over the baseline methods across all fairness metrics.

\begin{figure}[t]  
    \centering    
    \includegraphics[width=\linewidth]{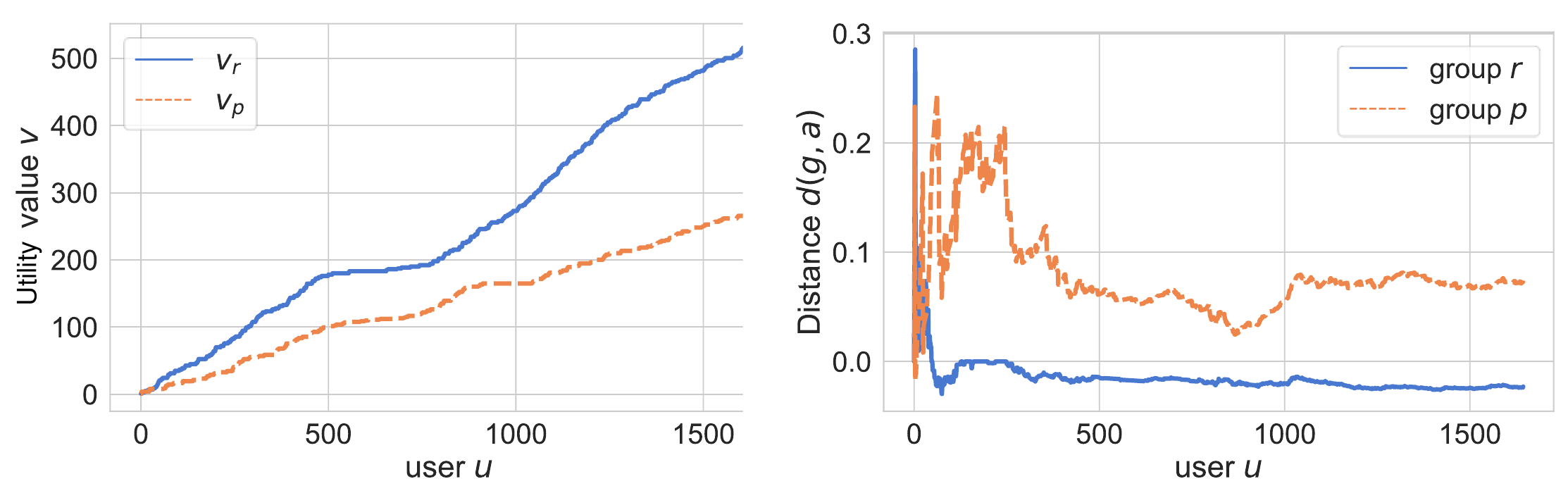}
    \caption{Group utility $\bm{v}_g$ and distance $d(g,a)$ \wrt the user $u$ arriving numbers, where $g$ is one rich group $r$ and poor group $p$; $\eta\%$ is set to $90\%$ and $t$ to $1.05$. }
    \label{fig:distance_exp} 
    \vspace{-0.3cm}
\end{figure}

\begin{figure}[t]  
    \centering    
    \includegraphics[width=\linewidth]{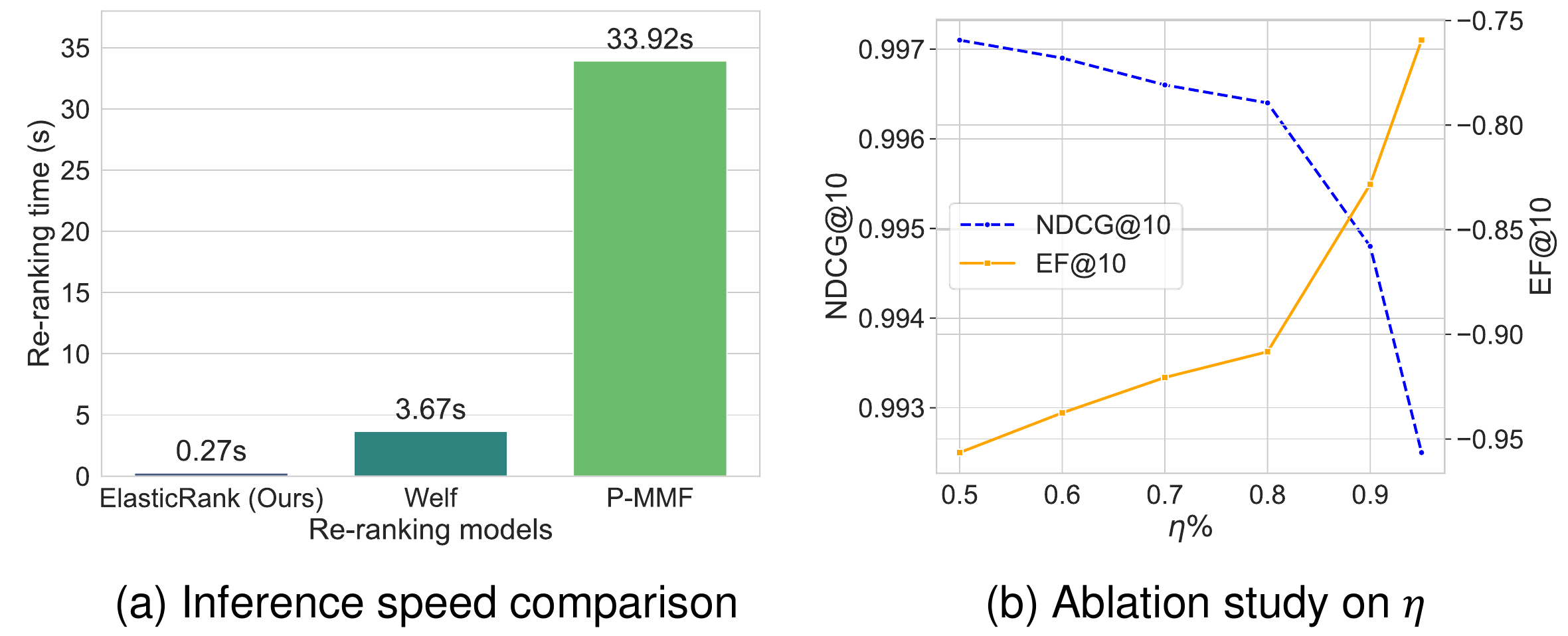}
    \caption{(a)~Inference speed comparison between ElasticRank and two best-performing baselines. (b)~NDCG@K and EF@K performance \wrt the anchor group radio $\eta\%$.}
    \label{fig:speed_ablation}  
    \vspace{-0.3cm}
\end{figure}

\subsection{Experimental analysis}
We also conduct experiments to analyze ElasticRank on the Steam dataset under $K=10$. Similar phenomena can be observed in the other two datasets and other $K$ values. 

\subsubsection{Visualization of distances between different groups}
In this section, we investigate how ElasticRank manages the accuracy-fairness trade-off by visualizing the curved distances between different groups.
Figure~\ref{fig:distance_exp} presents the group utility $\bm{v}_g$ and distance $d(g,a)$ \wrt the user $u$ arriving numbers, where $g$ is one rich group $r$ and poor group $p$. The experiments also are conducted on the Steam dataset with $K=10$. Intuitively, the distance $d(g,a)$ can be interpreted as a fairness score: smaller distances indicate that the system is less likely to provide exposure to the group, whereas larger distances suggest a higher likelihood of exposure for that group. The anchor radio $\eta\%$ is $90\%$.

From Figure~\ref{fig:distance_exp}, we can first observe that the rich group $r$ exhibits a higher utility than the poor group $p$, which helps ensure the accuracy performance of the system. However, to preserve fairness, the group 
$r$ is assigned smaller distances (fairness scores) relative to the anchor group $a$, indicating that ElasticRank is less likely to expose $r$ during the re-ranking process. Conversely, the group $p$ is assigned larger distances (fairness scores) relative to the anchor group $a$, highlighting ElasticRank's support for such a group.

\subsubsection{Inference speed}
In Figure~\ref{fig:speed_ablation} (a), we compare the inference time of the two best-performing baseline models, Welf and P-MMF (see performances in Figure~\ref{fig:Pareto}), with our model, ElasticRank, by testing the inference time across all users.

From Figure~\ref{fig:speed_ablation} (a), we can observe that ElasticRank is an order of magnitude faster than the Welf model in re-ranking and two orders of magnitude faster than P-MMF. This is because our model uses elasticity to dynamically compute fairness scores, rather than using optimization algorithms to compute gradients. This approach not only yields better results but also provides a significant improvement in efficiency, enabling rapid deployment in any industrial-level IR system. The computational complexity is the same as the standard sorting algorithms.

\subsubsection{Ablation study on choosing anchor group}
In this section, we conduct an ablation study for the anchor group setting radio $\eta\%$ (Eq.~(\ref{eq:curve_distance}) and
Figure~\ref{fig:speed_ablation} (b)) illustrates how the ranking accuracy (NDCG) and fairness metric (EF) \wrt $\eta\%$ from $50\%$-$95\%$.

From Figure ~\ref{fig:speed_ablation} (b), we found that $\eta$ can trade off accuracy and fairness, as increasing $\eta$ often improves fairness performance while reducing accuracy. Upon further investigation, we discovered that a larger anchor group ratio increases the gap between poorer groups and the anchor (richer) group, which results in more support for the poorer group to improve fairness. However, this approach tends to harm accuracy. Therefore, in practical applications, it is important to select different anchor group ratios based on specific needs to balance accuracy and fairness.
\section{Conclusion and Discussion}

\textbf{Conclusion.} In this paper, we understand accuracy-fairness trade-offs in re-ranking by framing them as a commodity taxation transfer problem. By leveraging elasticity theory from economics, we reveal that these trade-offs are determined by the elasticity between inter-groups. Inspired by elasticity theory, we introduce the EF-Curve, a evaluation framework for fair re-ranking, alongside ElasticRank, an algorithm that consistently outperforms state-of-the-art baselines in both efficiency and effectiveness. 

\textbf{Discussion.} Although we leverage the elasticity theory to analyze fairness in re-ranking, there are significant differences between them. In economics, the commodity tax transfer problem is inherently more complex because the tax revenue is often allocated to public goods, benefiting all users collectively. Additionally, savings mechanisms can restrict the impact of commodity taxes on users.
In contrast, re-ranking corresponds to a simpler, static setup, making it less intricate than its economic counterpart. In the future, the concept of commodity transfer could be extended to analyze and design more complex fair-aware IR applications.

\begin{acks}
This work was funded by the National Key R\&D Program of China (2023YFA1008704), the National Natural Science Foundation of China (No. 62472426, 62276248), the Youth Innovation Promotion Association CAS (No. 2023111). This research was (partially) supported by the Dutch Research Council (NWO), under project numbers 024.004.022, NWA.1389.20.\-183, and KICH3.LTP.20.006, and the European Union's Horizon Europe program under grant agreement No 101070212. All content represents the opinion of the authors, which is not necessarily shared or endorsed by their respective employers and/or sponsors.
\end{acks}
\appendix

\section*{Appendix}

\section{Proof of Theorem~\ref{theo:general_fair}}\label{app:proof_general_fair}

\begin{proof}
    First, we can use the normalized utility $\bar{\bm{v}}_g = \bm{v}_g/\sum_{g=1}^{|\mathcal{G}|}\bm{v}_g$ to make the function $f(\cdot)$ meet the second requirements scale-invariants: 
    $
        \bar{\bm{v}}_g = c\bm{v}_g/\sum_{g=1}^{|\mathcal{G}|}c\bm{v}_g.
    $

    Then, since $f(\cdot)$ should meet the third requirement, we can write:
    $
        f(\bm{1}_n/n) = f(\bm{1}_m/m), \forall n,m > 0
    $
    where the $\bm{1}_n$ means the $n$-th dimensional vector with all $1$ elements.
    Therefore, we can write
    \begin{equation}\label{eq:independent}
        \lim_{n \to \infty}\frac{f(\bm{1}_{n+1}/(n+1))}{f(\bm{1}_n/n)} = 1.
    \end{equation}

    Then, to meet the requirements for fairness and optimization objectives that should be distributed, we can make any data $\bm{x}\in \mathbb{R}^m$ partition into $[x_1, x_2, \cdots, x_n]$, where $n$ is the partition number. Then $f(x)$ can be aggregated as:
    $
        f(x) = h(f(x_1), f(x_2), \cdots, f(x_n)),
    $
    where $h$ is a Kolmogorov-Nagumo mean function~\cite{dukkipati2010kolmogorov} and  combining the requirements and the Equation~(\ref{eq:independent}) into the Theorem 2 of~\cite{lan2010axiomatic}, we can write $h$ as 
    $
        h = k^{-1}(\sum_{g=1}^{|\mathcal{G}|} k(\bar{\bm{v}}_g)),
    $
    where the function $k(\cdot)$ should only be the power family logarithm or power generator function: $k(x)=\log x$ or $k(x)=x^t$. 

    Therefore, the function is generated by the power function:
    \[
    f(\bm{v}; t) = \text{sign}(1-t)\left[\sum_{g=1}^{|\mathcal{G}|} \bar{\bm{v}}_g^{1-t}\right]^{\frac{1}{t}}.
    \]
    When $t\to 0$, the function $f$ becomes logarithm:
    \begin{align*}
        &\lim_{t\to 0} \log f(\bm{v}; t) = \lim_{t \to 0} \log\left[\sum_{g=1}^{|\mathcal{G}|} \bar{\bm{v}}_g^{1-t}\right]^{\frac{1}{t}}=\lim_{t\to 0} \frac{\left[\sum_{g=1}^{|\mathcal{G}|} \bar{\bm{v}}_g^{1-t}\right]}{t} \\
        &= \lim_{t\to 0} - \sum_{g=1}^{|\mathcal{G}|}(\log \bar{\bm{v}}_g) \bar{\bm{v}}_g^{1-t} = - \sum_{g=1}^{|\mathcal{G}|}\bar{\bm{v}}_g\log \bar{\bm{v}}_g = H(\bar{\bm{v}}),
    \end{align*}
    which will be reduced to the entropy fairness $H(\bm{v})$. Therefore, we have
    $
        \lim_{t \to 0}f(\bm{v}; t) = e^{H(\bar{\bm{v}})}.
    $
    When $t\to \infty$, the fairness formulation simplifies to the infinity norm, effectively reducing to max-min fairness, while other types of fairness can be easily mapped to their corresponding relationships.
    %
    %
    %
\end{proof}

\section{Proof of Theorem~\ref{theo:tax_t}}\label{app:proof_tax_t}

\begin{proof}
    According to elasticity definition in Equation~(\ref{eq:elastic}):
    \begin{equation}\label{eq:partial_f}
        \frac{\partial f(\bm{v};t)}{\partial \bar{\bm{v}}_g} = \frac{|1-t|}{t}\bar{\bm{v}}_g^{-t}\left[\sum_{g=1}^{|\mathcal{G}|} \bar{\bm{v}}_g^{1-t}\right]^{\frac{1}{t}-1}.
    \end{equation}
    Then, the elasticity can be measured through:
    \begin{align*}
        E_{r,p} = \frac{\partial v_p}{\partial v_r} = \frac{\frac{\partial f(\bm{v};t)}{\partial \bm{v}_r}}{\frac{\partial f(\bm{v};t)}{\partial \bm{v}_p}} = \left(\frac{\bar{\bm{v}}_r}{\bar{\bm{v}}_p}\right)^{-|t|},
    \end{align*}
   where the absolute value of $1-t$ means the fairness metric typically taxes the rich group and subsidizes the poor group to ensure proper redistribution. 
   \begin{equation}
       \frac{\partial f(\bm{v};t)}{\partial t}\text{sign}(1-t)\ge 0.
   \end{equation} 
   Since let $t_1>t_2>1$, $\phi(y) = y^{\frac{t_2}{t_1}}$ is concave, we have:

   
   \begin{align*}
       f(\bm{v};t_2) &= \ge -\left[\sum_{g=1}^{|\mathcal{G}|} \phi(\bar{\bm{v}}_g^{-t_1})\right]^{\frac{1}{t_2}}
       &\ge -\left[\phi(\sum_{g=1}^{|\mathcal{G}|} \bar{\bm{v}}_g^{-t_1})\right]^{\frac{1}{t_2}} \ge f(\bm{v};t_1),
   \end{align*}
   where the third step follows from Jensen’s inequality.
   
   Next, we will prove how to distinguish the poor and rich item groups. we check the $\frac{\partial f(\bm{v};t)}{\bm{v}_g}=0$ has a single root of
   $
   \bm{v}_g^{*} = \left(\frac{\sum_{g=1}^{|\mathcal{G}|}v_g}{\sum_{g=1}^{|\mathcal{G}|}v_g^{1-t}}\right)^{\frac{1}{t}},
   $
   where for $t\neq 1$, $\frac{\partial f(\bm{v};t)}{\partial \bm{v}_g} > 0$, if $\bm{v}_g > \theta$, otherwise, $\frac{\partial f(\bm{v};t)}{\partial \bm{v}_g} < 0$.
\end{proof}

\section{Proof of Theorem~\ref{theo:trade-off}}\label{app:proof_trade-off}

\begin{proof}
    
First, we  re-write the Equation~(\ref{eq:tradeoff_v1}) as
$L_1 = \sum_{g\in\mathcal{G}}\bm{v}_g + \lambda f(\bm{v})$.
Then we can write:
\begin{align*}
    L = \left[f(\bm{v};|t|)^{|t|} \cdot a(\bm{w})^{1-|t|}\right] \propto \sum_{g\in\mathcal{G}} \bm{v}_g^{1-|t|}.
\end{align*}
Let $l(\bm{v};r) = \sum_{g\in\mathcal{G}} \bm{v}_g^{1-r}$, $r\ge 0$, then 
$
    L_1 = l(\bm{v};0) + l(\bm{v}; |t|).
$
Since $l(\bm{v};r)$ is continuous \wrt $r$ and the feasible region of $\bm{v}$ is convex and continuous (because $\bm{v}$ is the linear transformation over a simplex space~\citep{lindenstrauss1978poulsen}), there exists a constant number $\lambda \ge 0$ s.t.
$
    \sum_{g\in\mathcal{G}}\bm{v}_g + \lambda f(\bm{v}; |t|) = L_1.
$
Therefore, the $\argmax_{\bm{v}} L_1 = \argmax_{\bm{v}} L$.

Then, we define the accuracy gradient $\alpha = \bm{1}$ and fairness gradient $\eta = \bm{1}/|\mathcal{G}|-\bar{\bm{v}}$. Then we can write:
\[
\langle \nabla_{\bm{v}} L,\alpha \rangle = \sum_{g\in\mathcal{G}}\bm{v}_g^{-|t|}, \langle \nabla_{\bm{v}} L,\eta \rangle = \sum_{g\in\mathcal{G}}\left(\bm{v}_g^{-|t|}(\bm{1}-\frac{\bm{v}_g}{\sum_g \bm{v}_g})\right).
\]
Therefore, 
\begin{align*}
    \gamma &= \frac{\langle \nabla_{\bm{v}} L,\eta \rangle}{\langle \nabla_{\bm{v}} L,\alpha \rangle} = 1-\frac{\sum_g\bar{\bm{v}}\bm{v}_g^{-|t|}}{\sum_g \bm{v}_g^{-|t|}}\\
    &= 1-\frac{\sum_g \bm{v}_g^{1-|t|}}{\sum_g \bm{v}_g \sum_{g}\bm{v}_g^{|t|}}=1-\frac{\sum_g \bm{v}_g^{1-|t|}}{\sum_g \bm{v}_g^{1-|t|}+\sum_{g}\sum_{r\neq g}\bm{v}_g^{-|t|} \bm{v}_r} \\
    &=1-\frac{1}{1+ \frac{\sum_{p\in\mathcal{G}}\sum_{r\neq p}\bm{v}_p^{1-|t|}E_{r,p}}{\sum_{p\in\mathcal{G}} \bm{v}_p^{1-|t|}}}=1-\frac{1}{1+k(E_{r,p})}.
    \qedhere
\end{align*}

\end{proof}

\newpage
\bibliographystyle{ACM-Reference-Format}
\bibliography{references}


\begin{thebibliography}{46}


\ifx \showCODEN    \undefined \def \showCODEN     #1{\unskip}     \fi
\ifx \showDOI      \undefined \def \showDOI       #1{#1}\fi
\ifx \showISBNx    \undefined \def \showISBNx     #1{\unskip}     \fi
\ifx \showISBNxiii \undefined \def \showISBNxiii  #1{\unskip}     \fi
\ifx \showISSN     \undefined \def \showISSN      #1{\unskip}     \fi
\ifx \showLCCN     \undefined \def \showLCCN      #1{\unskip}     \fi
\ifx \shownote     \undefined \def \shownote      #1{#1}          \fi
\ifx \showarticletitle \undefined \def \showarticletitle #1{#1}   \fi
\ifx \showURL      \undefined \def \showURL       {\relax}        \fi
\providecommand\bibfield[2]{#2}
\providecommand\bibinfo[2]{#2}
\providecommand\natexlab[1]{#1}
\providecommand\showeprint[2][]{arXiv:#2}

\bibitem[Abdollahpouri et~al\mbox{.}(2020)]%
        {abdollahpouri2020multistakeholder}
\bibfield{author}{\bibinfo{person}{Himan Abdollahpouri}, \bibinfo{person}{Gediminas Adomavicius}, \bibinfo{person}{Robin Burke}, \bibinfo{person}{Ido Guy}, \bibinfo{person}{Dietmar Jannach}, \bibinfo{person}{Toshihiro Kamishima}, \bibinfo{person}{Jan Krasnodebski}, {and} \bibinfo{person}{Luiz Pizzato}.} \bibinfo{year}{2020}\natexlab{}.
\newblock \showarticletitle{Multistakeholder Recommendation: Survey and Research Directions}.
\newblock \bibinfo{journal}{\emph{User Modeling and User-Adapted Interaction}} \bibinfo{volume}{30}, \bibinfo{number}{1} (\bibinfo{year}{2020}), \bibinfo{pages}{127--158}.
\newblock


\bibitem[Abdollahpouri and Burke(2019)]%
        {abdollahpouri2019multi}
\bibfield{author}{\bibinfo{person}{Himan Abdollahpouri} {and} \bibinfo{person}{Robin Burke}.} \bibinfo{year}{2019}\natexlab{}.
\newblock \showarticletitle{Multi-stakeholder Recommendation and its Connection to Multi-sided Fairness}.
\newblock \bibinfo{journal}{\emph{arXiv preprint arXiv:1907.13158}} (\bibinfo{year}{2019}).
\newblock


\bibitem[Abdollahpouri et~al\mbox{.}(2019)]%
        {abdollahpouri2019unfairness}
\bibfield{author}{\bibinfo{person}{Himan Abdollahpouri}, \bibinfo{person}{Masoud Mansoury}, \bibinfo{person}{Robin Burke}, {and} \bibinfo{person}{Bamshad Mobasher}.} \bibinfo{year}{2019}\natexlab{}.
\newblock \showarticletitle{The Unfairness of Popularity Bias in Recommendation}.
\newblock \bibinfo{journal}{\emph{arXiv preprint arXiv:1907.13286}} (\bibinfo{year}{2019}).
\newblock


\bibitem[Bekta{\c{s}} and Letchford(2020)]%
        {bektacs2020using}
\bibfield{author}{\bibinfo{person}{Tolga Bekta{\c{s}}} {and} \bibinfo{person}{Adam~N Letchford}.} \bibinfo{year}{2020}\natexlab{}.
\newblock \showarticletitle{Using $\ell$p-Norms for Fairness in Combinatorial Optimisation}.
\newblock \bibinfo{journal}{\emph{Computers \& Operations Research}}  \bibinfo{volume}{120} (\bibinfo{year}{2020}), \bibinfo{pages}{104975}.
\newblock


\bibitem[Biswas et~al\mbox{.}(2021)]%
        {fairrecplus}
\bibfield{author}{\bibinfo{person}{Arpita Biswas}, \bibinfo{person}{Gourab~K. Patro}, \bibinfo{person}{Niloy Ganguly}, \bibinfo{person}{Krishna~P Gummadi}, {and} \bibinfo{person}{Abhijnan Chakraborty}.} \bibinfo{year}{2021}\natexlab{}.
\newblock \showarticletitle{Toward Fair Recommendation in Two-sided Platforms}.
\newblock \bibinfo{journal}{\emph{ACM Transactions on the Web (TWEB)}} \bibinfo{volume}{16}, \bibinfo{number}{2} (\bibinfo{year}{2021}), \bibinfo{pages}{1--34}.
\newblock


\bibitem[Bouttier et~al\mbox{.}(2003)]%
        {bouttier2003geodesic}
\bibfield{author}{\bibinfo{person}{J{\'e}r{\'e}mie Bouttier}, \bibinfo{person}{Philippe Di~Francesco}, {and} \bibinfo{person}{Emmanuel Guitter}.} \bibinfo{year}{2003}\natexlab{}.
\newblock \showarticletitle{Geodesic Distance in Planar Graphs}.
\newblock \bibinfo{journal}{\emph{Nuclear physics B}} \bibinfo{volume}{663}, \bibinfo{number}{3} (\bibinfo{year}{2003}), \bibinfo{pages}{535--567}.
\newblock


\bibitem[Calmon et~al\mbox{.}(2017)]%
        {Calmon17}
\bibfield{author}{\bibinfo{person}{Flavio~P. Calmon}, \bibinfo{person}{Dennis Wei}, \bibinfo{person}{Bhanukiran Vinzamuri}, \bibinfo{person}{Karthikeyan~Natesan Ramamurthy}, {and} \bibinfo{person}{Kush~R. Varshney}.} \bibinfo{year}{2017}\natexlab{}.
\newblock \showarticletitle{Optimized Pre-processing for Discrimination Prevention}. In \bibinfo{booktitle}{\emph{Proceedings of the 31st International Conference on Neural Information Processing Systems}} (Long Beach, California, USA) \emph{(\bibinfo{series}{NIPS'17})}. \bibinfo{publisher}{Curran Associates Inc.}, \bibinfo{address}{Red Hook, NY, USA}, \bibinfo{pages}{3995–4004}.
\newblock
\showISBNx{9781510860964}


\bibitem[Deldjoo et~al\mbox{.}(2019)]%
        {deldjoo2019recommender}
\bibfield{author}{\bibinfo{person}{Yashar Deldjoo}, \bibinfo{person}{Vito~Walter Anelli}, \bibinfo{person}{Hamed Zamani}, \bibinfo{person}{Alejandro Bellog{\'\i}n}, {and} \bibinfo{person}{Tommaso Di~Noia}.} \bibinfo{year}{2019}\natexlab{}.
\newblock \showarticletitle{Recommender Systems Fairness Evaluation via Generalized Cross Entropy}.
\newblock \bibinfo{journal}{\emph{arXiv preprint arXiv:1908.06708}} (\bibinfo{year}{2019}).
\newblock


\bibitem[Deldjoo et~al\mbox{.}(2022)]%
        {deldjoo2022survey}
\bibfield{author}{\bibinfo{person}{Yashar Deldjoo}, \bibinfo{person}{Dietmar Jannach}, \bibinfo{person}{Alejandro Bellogin}, \bibinfo{person}{Alessandro Difonzo}, {and} \bibinfo{person}{Dario Zanzonelli}.} \bibinfo{year}{2022}\natexlab{}.
\newblock \showarticletitle{A Survey of Research on Fair Recommender Systems}.
\newblock \bibinfo{journal}{\emph{arXiv preprint arXiv:2205.11127}} (\bibinfo{year}{2022}).
\newblock


\bibitem[Do et~al\mbox{.}(2021)]%
        {nips21welf}
\bibfield{author}{\bibinfo{person}{Virginie Do}, \bibinfo{person}{Sam Corbett-Davies}, \bibinfo{person}{Jamal Atif}, {and} \bibinfo{person}{Nicolas Usunier}.} \bibinfo{year}{2021}\natexlab{}.
\newblock \showarticletitle{Two-sided Fairness in Rankings via Lorenz Dominance}.
\newblock \bibinfo{journal}{\emph{Advances in Neural Information Processing Systems}}  \bibinfo{volume}{34} (\bibinfo{year}{2021}), \bibinfo{pages}{8596--8608}.
\newblock


\bibitem[Do and Usunier(2022)]%
        {do2022optimizing}
\bibfield{author}{\bibinfo{person}{Virginie Do} {and} \bibinfo{person}{Nicolas Usunier}.} \bibinfo{year}{2022}\natexlab{}.
\newblock \showarticletitle{Optimizing Generalized Gini Indices for Fairness in Rankings}. In \bibinfo{booktitle}{\emph{Proceedings of the 45th International ACM SIGIR Conference on Research and Development in Information Retrieval}}. \bibinfo{pages}{737--747}.
\newblock


\bibitem[Dukkipati(2010)]%
        {dukkipati2010kolmogorov}
\bibfield{author}{\bibinfo{person}{Ambedkar Dukkipati}.} \bibinfo{year}{2010}\natexlab{}.
\newblock \showarticletitle{On Kolmogorov-Nagumo Averages and Nonextensive Entropy}. In \bibinfo{booktitle}{\emph{2010 International Symposium On Information Theory \& Its Applications}}. IEEE, \bibinfo{pages}{446--451}.
\newblock


\bibitem[Hanlon and Heitzman(2010)]%
        {hanlon2010review}
\bibfield{author}{\bibinfo{person}{Michelle Hanlon} {and} \bibinfo{person}{Shane Heitzman}.} \bibinfo{year}{2010}\natexlab{}.
\newblock \showarticletitle{A Review of Tax Research}.
\newblock \bibinfo{journal}{\emph{Journal of accounting and Economics}} \bibinfo{volume}{50}, \bibinfo{number}{2-3} (\bibinfo{year}{2010}), \bibinfo{pages}{127--178}.
\newblock


\bibitem[He and McAuley(2016)]%
        {he2016ups}
\bibfield{author}{\bibinfo{person}{Ruining He} {and} \bibinfo{person}{Julian McAuley}.} \bibinfo{year}{2016}\natexlab{}.
\newblock \showarticletitle{Ups and downs: Modeling the visual evolution of fashion trends with one-class collaborative filtering}. In \bibinfo{booktitle}{\emph{proceedings of the 25th international conference on world wide web}}. \bibinfo{pages}{507--517}.
\newblock


\bibitem[Jaenich et~al\mbox{.}(2024)]%
        {jaenich2024fairness}
\bibfield{author}{\bibinfo{person}{Thomas Jaenich}, \bibinfo{person}{Graham McDonald}, {and} \bibinfo{person}{Iadh Ounis}.} \bibinfo{year}{2024}\natexlab{}.
\newblock \showarticletitle{Fairness-Aware Exposure Allocation via Adaptive Reranking}. In \bibinfo{booktitle}{\emph{Proceedings of the 47th International ACM SIGIR Conference on Research and Development in Information Retrieval}}. \bibinfo{pages}{1504--1513}.
\newblock


\bibitem[Kang and McAuley(2018)]%
        {SASRec}
\bibfield{author}{\bibinfo{person}{Wang-Cheng Kang} {and} \bibinfo{person}{Julian McAuley}.} \bibinfo{year}{2018}\natexlab{}.
\newblock \showarticletitle{Self-attentive Sequential Recommendation}. In \bibinfo{booktitle}{\emph{2018 IEEE international conference on data mining (ICDM)}}. IEEE, \bibinfo{pages}{197--206}.
\newblock


\bibitem[Lambert(1992)]%
        {lambert1992distribution}
\bibfield{author}{\bibinfo{person}{Peter~J. Lambert}.} \bibinfo{year}{1992}\natexlab{}.
\newblock \bibinfo{booktitle}{\emph{The Distribution and Redistribution of Income}}.
\newblock \bibinfo{publisher}{Springer}.
\newblock


\bibitem[Lan et~al\mbox{.}(2010)]%
        {lan2010axiomatic}
\bibfield{author}{\bibinfo{person}{Tian Lan}, \bibinfo{person}{David Kao}, \bibinfo{person}{Mung Chiang}, {and} \bibinfo{person}{Ashutosh Sabharwal}.} \bibinfo{year}{2010}\natexlab{}.
\newblock \showarticletitle{An Axiomatic Theory of Fairness in Network Resource Allocation}. In \bibinfo{booktitle}{\emph{Proceedings IEEE INFOCOM}}. \bibinfo{pages}{1--9}.
\newblock


\bibitem[Li et~al\mbox{.}(2021)]%
        {li2021user}
\bibfield{author}{\bibinfo{person}{Yunqi Li}, \bibinfo{person}{Hanxiong Chen}, \bibinfo{person}{Zuohui Fu}, \bibinfo{person}{Yingqiang Ge}, {and} \bibinfo{person}{Yongfeng Zhang}.} \bibinfo{year}{2021}\natexlab{}.
\newblock \showarticletitle{User-oriented Fairness in Recommendation}. In \bibinfo{booktitle}{\emph{Proceedings of the Web Conference}}. \bibinfo{pages}{624--632}.
\newblock


\bibitem[Li et~al\mbox{.}(2022)]%
        {li2022fairness}
\bibfield{author}{\bibinfo{person}{Yunqi Li}, \bibinfo{person}{Hanxiong Chen}, \bibinfo{person}{Shuyuan Xu}, \bibinfo{person}{Yingqiang Ge}, \bibinfo{person}{Juntao Tan}, \bibinfo{person}{Shuchang Liu}, {and} \bibinfo{person}{Yongfeng Zhang}.} \bibinfo{year}{2022}\natexlab{}.
\newblock \showarticletitle{Fairness in Recommendation: A Survey}.
\newblock \bibinfo{journal}{\emph{arXiv preprint arXiv:2205.13619}} (\bibinfo{year}{2022}).
\newblock


\bibitem[Li et~al\mbox{.}(2023)]%
        {lifairness}
\bibfield{author}{\bibinfo{person}{Yunqi Li}, \bibinfo{person}{Hanxiong Chen}, \bibinfo{person}{Shuyuan Xu}, \bibinfo{person}{Yingqiang Ge}, \bibinfo{person}{Juntao Tan}, \bibinfo{person}{Shuchang Liu}, {and} \bibinfo{person}{Yongfeng Zhang}.} \bibinfo{year}{2023}\natexlab{}.
\newblock \showarticletitle{Fairness in Recommendation: Foundations, Methods and Applications}.
\newblock \bibinfo{journal}{\emph{ACM Transactions on Intelligent Systems and Technology}} \bibinfo{volume}{14}, \bibinfo{number}{5} (\bibinfo{year}{2023}), \bibinfo{pages}{Article No.: 95}.
\newblock


\bibitem[Lindenstrauss et~al\mbox{.}(1978)]%
        {lindenstrauss1978poulsen}
\bibfield{author}{\bibinfo{person}{Joram Lindenstrauss}, \bibinfo{person}{Gunnar Olsen}, {and} \bibinfo{person}{Yaki Sternfeld}.} \bibinfo{year}{1978}\natexlab{}.
\newblock \showarticletitle{The Poulsen Simplex}. In \bibinfo{booktitle}{\emph{Annales de l'Institut Fourier}}, Vol.~\bibinfo{volume}{28}. \bibinfo{pages}{91--114}.
\newblock


\bibitem[Lipani(2016)]%
        {lipani2016fairness}
\bibfield{author}{\bibinfo{person}{Aldo Lipani}.} \bibinfo{year}{2016}\natexlab{}.
\newblock \showarticletitle{Fairness in Information Retrieval}. In \bibinfo{booktitle}{\emph{Proceedings of the 39th International ACM SIGIR conference on Research and Development in Information Retrieval}}. \bibinfo{pages}{1171--1171}.
\newblock


\bibitem[Liu et~al\mbox{.}(2021)]%
        {liu2021neural}
\bibfield{author}{\bibinfo{person}{Xiangyu Liu}, \bibinfo{person}{Chuan Yu}, \bibinfo{person}{Zhilin Zhang}, \bibinfo{person}{Zhenzhe Zheng}, \bibinfo{person}{Yu Rong}, \bibinfo{person}{Hongtao Lv}, \bibinfo{person}{Da Huo}, \bibinfo{person}{Yiqing Wang}, \bibinfo{person}{Dagui Chen}, \bibinfo{person}{Jian Xu}, {et~al\mbox{.}}} \bibinfo{year}{2021}\natexlab{}.
\newblock \showarticletitle{Neural Auction: End-to-end Learning of Auction Mechanisms for E-commerce Advertising}. In \bibinfo{booktitle}{\emph{Proceedings of the 27th ACM Conference on Knowledge Discovery \& Data Mining}}. \bibinfo{pages}{3354--3364}.
\newblock


\bibitem[Lotov and Miettinen(2008)]%
        {lotov2008visualizing}
\bibfield{author}{\bibinfo{person}{Alexander~V Lotov} {and} \bibinfo{person}{Kaisa Miettinen}.} \bibinfo{year}{2008}\natexlab{}.
\newblock \showarticletitle{Visualizing the Pareto frontier}.
\newblock In \bibinfo{booktitle}{\emph{Multiobjective optimization}}. \bibinfo{publisher}{Springer}, \bibinfo{pages}{213--243}.
\newblock


\bibitem[Naghiaei et~al\mbox{.}(2022)]%
        {cpfair}
\bibfield{author}{\bibinfo{person}{Mohammadmehdi Naghiaei}, \bibinfo{person}{Hossein~A Rahmani}, {and} \bibinfo{person}{Yashar Deldjoo}.} \bibinfo{year}{2022}\natexlab{}.
\newblock \showarticletitle{CPFair: Personalized Consumer and Producer Fairness Re-ranking for Recommender Systems}.
\newblock \bibinfo{journal}{\emph{arXiv preprint arXiv:2204.08085}} (\bibinfo{year}{2022}).
\newblock


\bibitem[Nerr{\'e}(2001)]%
        {nerre2001concept}
\bibfield{author}{\bibinfo{person}{Birger Nerr{\'e}}.} \bibinfo{year}{2001}\natexlab{}.
\newblock \showarticletitle{The Concept of Tax Culture}. In \bibinfo{booktitle}{\emph{Proceedings Annual Conference on Taxation and Minutes of the Annual Meeting of the National Tax Association}}, Vol.~\bibinfo{volume}{94}. JSTOR, \bibinfo{pages}{288--295}.
\newblock


\bibitem[Patro et~al\mbox{.}(2020)]%
        {fairrec}
\bibfield{author}{\bibinfo{person}{Gourab~K. Patro}, \bibinfo{person}{Arpita Biswas}, \bibinfo{person}{Niloy Ganguly}, \bibinfo{person}{Krishna~P. Gummadi}, {and} \bibinfo{person}{Abhijnan Chakraborty}.} \bibinfo{year}{2020}\natexlab{}.
\newblock \showarticletitle{FairRec: Two-sided Fairness for Personalized Recommendations in Two-sided Platforms}. In \bibinfo{booktitle}{\emph{Proceedings of the Web Conference}}. \bibinfo{pages}{1194--1204}.
\newblock


\bibitem[Ramsey(1927)]%
        {ramsey1927contribution}
\bibfield{author}{\bibinfo{person}{Frank~P. Ramsey}.} \bibinfo{year}{1927}\natexlab{}.
\newblock \showarticletitle{A Contribution to the Theory of Taxation}.
\newblock \bibinfo{journal}{\emph{The Economic Journal}} \bibinfo{volume}{37}, \bibinfo{number}{145} (\bibinfo{year}{1927}), \bibinfo{pages}{47--61}.
\newblock


\bibitem[Renner and Wolf(2004)]%
        {renner2004smooth}
\bibfield{author}{\bibinfo{person}{Renato Renner} {and} \bibinfo{person}{Stefan Wolf}.} \bibinfo{year}{2004}\natexlab{}.
\newblock \showarticletitle{Smooth R{\'e}nyi Entropy and Applications}. In \bibinfo{booktitle}{\emph{Proceedings International Symposium on Information Theory. ISIT 2004}}. IEEE, \bibinfo{pages}{233}.
\newblock


\bibitem[Robbins(1997)]%
        {robbins1997elasticity}
\bibfield{author}{\bibinfo{person}{Lionel Robbins}.} \bibinfo{year}{1997}\natexlab{}.
\newblock \showarticletitle{On the Elasticity of Demand for Income in Terms of Effort}.
\newblock In \bibinfo{booktitle}{\emph{Economic Science and Political Economy: Selected Articles}}. \bibinfo{publisher}{Springer}, \bibinfo{pages}{79--84}.
\newblock


\bibitem[Saito and Joachims(2022)]%
        {saito2022fair}
\bibfield{author}{\bibinfo{person}{Yuta Saito} {and} \bibinfo{person}{Thorsten Joachims}.} \bibinfo{year}{2022}\natexlab{}.
\newblock \showarticletitle{Fair Ranking as Fair Division: Impact-Based Individual Fairness in Ranking}. In \bibinfo{booktitle}{\emph{Proceedings of the 28th ACM SIGKDD Conference on Knowledge Discovery and Data Mining}}. \bibinfo{pages}{1514--1524}.
\newblock


\bibitem[Singh and Joachims(2019)]%
        {singh2019policy}
\bibfield{author}{\bibinfo{person}{Ashudeep Singh} {and} \bibinfo{person}{Thorsten Joachims}.} \bibinfo{year}{2019}\natexlab{}.
\newblock \showarticletitle{Policy Learning for Fairness in Ranking}.
\newblock \bibinfo{journal}{\emph{Advances in Neural Information Processing Systems}}  \bibinfo{volume}{32} (\bibinfo{year}{2019}).
\newblock


\bibitem[Tang et~al\mbox{.}(2023)]%
        {Tang23FairBias}
\bibfield{author}{\bibinfo{person}{Jiakai Tang}, \bibinfo{person}{Shiqi Shen}, \bibinfo{person}{Zhipeng Wang}, \bibinfo{person}{Zhi Gong}, \bibinfo{person}{Jingsen Zhang}, {and} \bibinfo{person}{Xu Chen}.} \bibinfo{year}{2023}\natexlab{}.
\newblock \showarticletitle{When Fairness meets Bias: a Debiased Framework for Fairness aware Top-N Recommendation}. In \bibinfo{booktitle}{\emph{Proceedings of the 17th ACM Conference on Recommender Systems}} (Singapore, Singapore) \emph{(\bibinfo{series}{RecSys '23})}. \bibinfo{publisher}{Association for Computing Machinery}, \bibinfo{address}{New York, NY, USA}, \bibinfo{pages}{200–210}.
\newblock
\showISBNx{9798400702419}
\urldef\tempurl%
\url{https://doi.org/10.1145/3604915.3608770}
\showDOI{\tempurl}


\bibitem[Wei et~al\mbox{.}(2022)]%
        {wei2022rank}
\bibfield{author}{\bibinfo{person}{Dong Wei}, \bibinfo{person}{Md~Mouinul Islam}, \bibinfo{person}{Baruch Schieber}, {and} \bibinfo{person}{Senjuti Basu~Roy}.} \bibinfo{year}{2022}\natexlab{}.
\newblock \showarticletitle{Rank Aggregation with Proportionate Fairness}. In \bibinfo{booktitle}{\emph{Proceedings of the 2022 international conference on management of data}}. \bibinfo{pages}{262--275}.
\newblock


\bibitem[Wu et~al\mbox{.}(2021)]%
        {wu2021tfrom}
\bibfield{author}{\bibinfo{person}{Yao Wu}, \bibinfo{person}{Jian Cao}, \bibinfo{person}{Guandong Xu}, {and} \bibinfo{person}{Yudong Tan}.} \bibinfo{year}{2021}\natexlab{}.
\newblock \showarticletitle{TFROM: A Two-sided Fairness-aware Recommendation Model for both Customers and Providers}. In \bibinfo{booktitle}{\emph{Proceedings of the 44th International ACM SIGIR Conference on Research and Development in Information Retrieval}}. \bibinfo{pages}{1013--1022}.
\newblock


\bibitem[Xiong et~al\mbox{.}(2024)]%
        {xiong2024fairwasp}
\bibfield{author}{\bibinfo{person}{Zikai Xiong}, \bibinfo{person}{Niccol{\`o} Dalmasso}, \bibinfo{person}{Alan Mishler}, \bibinfo{person}{Vamsi~K Potluru}, \bibinfo{person}{Tucker Balch}, {and} \bibinfo{person}{Manuela Veloso}.} \bibinfo{year}{2024}\natexlab{}.
\newblock \showarticletitle{Fairwasp: Fast and optimal fair wasserstein pre-processing}. In \bibinfo{booktitle}{\emph{Proceedings of the AAAI Conference on Artificial Intelligence}}, Vol.~\bibinfo{volume}{38}. \bibinfo{pages}{16120--16128}.
\newblock


\bibitem[Xu et~al\mbox{.}(2023)]%
        {xu2023p}
\bibfield{author}{\bibinfo{person}{Chen Xu}, \bibinfo{person}{Sirui Chen}, \bibinfo{person}{Jun Xu}, \bibinfo{person}{Weiran Shen}, \bibinfo{person}{Xiao Zhang}, \bibinfo{person}{Gang Wang}, {and} \bibinfo{person}{Zhenhua Dong}.} \bibinfo{year}{2023}\natexlab{}.
\newblock \showarticletitle{P-MMF: Provider Max-min Fairness Re-ranking in Recommender System}. In \bibinfo{booktitle}{\emph{Proceedings of the ACM Web Conference 2023}}. \bibinfo{pages}{3701--3711}.
\newblock


\bibitem[Xu et~al\mbox{.}(2025)]%
        {xu2025fairdiversecomprehensivetoolkitfair}
\bibfield{author}{\bibinfo{person}{Chen Xu}, \bibinfo{person}{Zhirui Deng}, \bibinfo{person}{Clara Rus}, \bibinfo{person}{Xiaopeng Ye}, \bibinfo{person}{Yuanna Liu}, \bibinfo{person}{Jun Xu}, \bibinfo{person}{Zhicheng Dou}, \bibinfo{person}{Ji-Rong Wen}, {and} \bibinfo{person}{Maarten de Rijke}.} \bibinfo{year}{2025}\natexlab{}.
\newblock \bibinfo{title}{FairDiverse: A Comprehensive Toolkit for Fair and Diverse Information Retrieval Algorithms}.
\newblock
\newblock
\showeprint[arxiv]{2502.11883}~[cs.IR]
\urldef\tempurl%
\url{https://arxiv.org/abs/2502.11883}
\showURL{%
\tempurl}


\bibitem[Xu et~al\mbox{.}(2024a)]%
        {fairsync}
\bibfield{author}{\bibinfo{person}{Chen Xu}, \bibinfo{person}{Jun Xu}, \bibinfo{person}{Yiming Ding}, \bibinfo{person}{Xiao Zhang}, {and} \bibinfo{person}{Qi Qi}.} \bibinfo{year}{2024}\natexlab{a}.
\newblock \showarticletitle{FairSync: Ensuring Amortized Group Exposure in Distributed Recommendation Retrieval}. In \bibinfo{booktitle}{\emph{Proceedings of the ACM Web Conference 2024}} (Singapore, Singapore) \emph{(\bibinfo{series}{WWW '24})}. \bibinfo{publisher}{Association for Computing Machinery}, \bibinfo{address}{New York, NY, USA}, \bibinfo{pages}{1092–1102}.
\newblock
\showISBNx{9798400701719}
\urldef\tempurl%
\url{https://doi.org/10.1145/3589334.3645413}
\showDOI{\tempurl}


\bibitem[Xu et~al\mbox{.}(2024b)]%
        {TaxRank}
\bibfield{author}{\bibinfo{person}{Chen Xu}, \bibinfo{person}{Xiaopeng Ye}, \bibinfo{person}{Wenjie Wang}, \bibinfo{person}{Liang Pang}, \bibinfo{person}{Jun Xu}, {and} \bibinfo{person}{Tat-Seng Chua}.} \bibinfo{year}{2024}\natexlab{b}.
\newblock \showarticletitle{A Taxation Perspective for Fair Re-ranking}. In \bibinfo{booktitle}{\emph{Proceedings of the 47th International ACM SIGIR Conference on Research and Development in Information Retrieval}} (Washington DC, USA) \emph{(\bibinfo{series}{SIGIR '24})}. \bibinfo{publisher}{Association for Computing Machinery}, \bibinfo{address}{New York, NY, USA}, \bibinfo{pages}{1494–1503}.
\newblock
\showISBNx{9798400704314}
\urldef\tempurl%
\url{https://doi.org/10.1145/3626772.3657766}
\showDOI{\tempurl}


\bibitem[Xue et~al\mbox{.}(2017)]%
        {DMF}
\bibfield{author}{\bibinfo{person}{Hong-Jian Xue}, \bibinfo{person}{Xinyu Dai}, \bibinfo{person}{Jianbing Zhang}, \bibinfo{person}{Shujian Huang}, {and} \bibinfo{person}{Jiajun Chen}.} \bibinfo{year}{2017}\natexlab{}.
\newblock \showarticletitle{Deep Matrix Factorization Models for Recommender Systems.}. In \bibinfo{booktitle}{\emph{IJCAI}}, Vol.~\bibinfo{volume}{17}. Melbourne, Australia, \bibinfo{pages}{3203--3209}.
\newblock


\bibitem[Yang et~al\mbox{.}(2023)]%
        {TaoSIGIRAP}
\bibfield{author}{\bibinfo{person}{Tao Yang}, \bibinfo{person}{Zhichao Xu}, {and} \bibinfo{person}{Qingyao Ai}.} \bibinfo{year}{2023}\natexlab{}.
\newblock \showarticletitle{Vertical Allocation-based Fair Exposure Amortizing in Ranking}. In \bibinfo{booktitle}{\emph{Proceedings of the Annual International ACM SIGIR Conference on Research and Development in Information Retrieval in the Asia Pacific Region}} (Beijing, China) \emph{(\bibinfo{series}{SIGIR-AP '23})}. \bibinfo{publisher}{Association for Computing Machinery}, \bibinfo{address}{New York, NY, USA}, \bibinfo{pages}{234–244}.
\newblock
\showISBNx{9798400704086}
\urldef\tempurl%
\url{https://doi.org/10.1145/3624918.3625313}
\showDOI{\tempurl}


\bibitem[Yang et~al\mbox{.}(2019)]%
        {yang2019bid}
\bibfield{author}{\bibinfo{person}{Xun Yang}, \bibinfo{person}{Yasong Li}, \bibinfo{person}{Hao Wang}, \bibinfo{person}{Di Wu}, \bibinfo{person}{Qing Tan}, \bibinfo{person}{Jian Xu}, {and} \bibinfo{person}{Kun Gai}.} \bibinfo{year}{2019}\natexlab{}.
\newblock \showarticletitle{Bid Optimization by Multivariable Control in Display Advertising}. In \bibinfo{booktitle}{\emph{Proceedings of the 25th ACM international conference on knowledge discovery \& data mining}}. \bibinfo{pages}{1966--1974}.
\newblock


\bibitem[Ye et~al\mbox{.}(2024)]%
        {BankFair}
\bibfield{author}{\bibinfo{person}{Xiaopeng Ye}, \bibinfo{person}{Chen Xu}, \bibinfo{person}{Jun Xu}, \bibinfo{person}{Xuyang Xie}, \bibinfo{person}{Gang Wang}, {and} \bibinfo{person}{Zhenhua Dong}.} \bibinfo{year}{2024}\natexlab{}.
\newblock \showarticletitle{Guaranteeing Accuracy and Fairness under Fluctuating User Traffic: A Bankruptcy-Inspired Re-ranking Approach} \emph{(\bibinfo{series}{CIKM '24})}. \bibinfo{publisher}{Association for Computing Machinery}, \bibinfo{address}{New York, NY, USA}, \bibinfo{pages}{2991–3001}.
\newblock
\showISBNx{9798400704369}
\urldef\tempurl%
\url{https://doi.org/10.1145/3627673.3679590}
\showDOI{\tempurl}


\bibitem[Zafar et~al\mbox{.}(2019)]%
        {zafar2019fairness}
\bibfield{author}{\bibinfo{person}{Muhammad~Bilal Zafar}, \bibinfo{person}{Isabel Valera}, \bibinfo{person}{Manuel Gomez-Rodriguez}, {and} \bibinfo{person}{Krishna~P Gummadi}.} \bibinfo{year}{2019}\natexlab{}.
\newblock \showarticletitle{Fairness Constraints: A Flexible Approach for Fair Classification}.
\newblock \bibinfo{journal}{\emph{The Journal of Machine Learning Research}} \bibinfo{volume}{20}, \bibinfo{number}{1} (\bibinfo{year}{2019}), \bibinfo{pages}{2737--2778}.
\newblock


\end{thebibliography}

\end{document}